\documentclass[10pt,letterpaper]{article}
\usepackage[paperwidth=171mm,paperheight=246mm,
  left=14.3mm,textwidth=142.4mm,lines=47,
  headsep=\baselineskip,headheight=16mm,
  footskip=23mm]{geometry}

\usepackage{color}
\usepackage{ulem}
\usepackage{natbib}
\newcommand{\delete}[1]{}

\newcommand{\deleted}[1]{}

\usepackage{siunitx}
\usepackage{hhline}
\usepackage{multicol}
\usepackage{url,booktabs,multirow}
\usepackage{caption}
\usepackage{subcaption}
\usepackage{color}
\usepackage{enumitem}
\usepackage{hyperref,amsmath}
\usepackage{algorithm2e}
\usepackage{comment}
\usepackage{mathtools}
\usepackage{footnote}

\usepackage[bb=dsserif]{mathalpha}
\usepackage{bm}

\usepackage[T1]{fontenc}
\usepackage[utf8]{inputenc}
\usepackage[default]{lato}

\usepackage{scrextend}
\changefontsizes[11]{7.5}

\usepackage{hyperref}

\RestyleAlgo{ruled}
\SetKwComment{Comment}{/* }{ */}

\begin{document}

\begin{center}
{\Large\textbf\newline{Unveiling Local Patterns of Child Pornography Consumption in France using Tor}
}
\end{center}

\begin{center}
{Till Koebe\textsuperscript{1*}, Zinnya del Villar\textsuperscript{2}, Brahmani Nutakki\textsuperscript{1}, Nursulu Sagimbayeva\textsuperscript{1}, Ingmar Weber\textsuperscript{1}
}
\end{center}

\noindent\textbf{1} Saarland Informatics Campus, Saarland University, Saarbrücken, Germany\\
\noindent\textbf{2} Data-Pop Alliance, Rennes, France
\\

\noindent\textbf{*} Corresponding author: till.koebe@uni-saarland.de
\\


\begin{abstract}

Child pornography represents a severe form of exploitation and victimization of children, leaving the victims with emotional and physical trauma. In this study, we aim to analyze local patterns of child pornography consumption across 1341 French communes in 20 metropolitan regions of France using fine-grained mobile traffic data of Tor network-related web services. We estimate that approx. 0.08 \% of Tor mobile download traffic observed in France is linked to the consumption of child sexual abuse materials by correlating it with local-level temporal porn consumption patterns. This compares to 0.19 \% of what we conservatively estimate to be the share of child pornographic content in global Tor traffic. In line with existing literature on the link between sexual child abuse and the consumption of image-based content thereof, we observe a positive and statistically significant effect of our child pornography consumption estimates on the reported number of victims of sexual violence and vice versa, which validates our findings, after controlling for a set of spatial and non-spatial features including socio-demographic characteristics, voting behaviour, nearby points of interest and Google Trends queries. While this is a first, exploratory attempt to look at child pornography from a spatial epidemiological angle, we believe this research provides public health officials with valuable information to prioritize target areas for public awareness campaigns as another step to fulfill the global community's pledge to target 16.2 of the Sustainable Development Goals: ``End abuse, exploitation, trafficking and all forms of violence and torture against children".

\end{abstract}

\section{Introduction}\label{section-intro}

\begin{center}
{"\textit{Derrière tout échange d’image ou de vidéo pédopornographique, il y a un agresseur et un mineur agressé.}" -- Adrien Taquet, 2021.

(Behind any exchange of child pornographic images or videos, there is an attacker and an attacked minor.)}
\end{center}

As pointed out by the French Secretary of State for Child Protection Adrien Taquet in 2021, child sexual abuse materials (CSAM), better known as child pornography, represent both a severe form of exploitation and victimization of children and at the same time a criminal offense \cite{taquet2021}. Sexual violence leaves affected children with emotional and physical trauma \cite{pinheiro2006violence}. For France, the National Institute of Health and Medical Research (INSERM) estimated in a general population survey, conducted between 2020 and 2021 that 1 in 10 French adults, approx. 5.5 million individuals, have been subject to sexual violence in their childhood \cite{sauve2021violences}, with serious health consequences as shown by \citet{brown2023introduction}. The Independent Commission on Incest and Sexual Violence against Children (CIIVISE) installed by the French president on March 23, 2021, estimates that every year in France alone 160,000 children become victims of sexual violence. \citet{eke_examining_2011} found that 24\% of child pornography users from their sample had committed sexual offenses in the past. Similarly, \citet{Hall2007APO} reported that 30\% to 80\% of individuals who viewed child pornography had molested a child. That emphasizes the important link between CSAM consumption and sexual violence against children.

\begin{center}
    \noindent\fbox{%
    \parbox{0.98\textwidth}{%
        \textbf{Descriptions of key terms used in this study }
        \begin{itemize}
            \item \textbf{Tor:} The Tor network, short for ``The Onion Router," is a privacy-focused network that directs internet traffic through a series of volunteer-operated servers, encrypting it at each step and making it difficult to trace back to the user's origin, thereby enhancing online anonymity. It is commonly used to access the internet anonymously.
            \item \textbf{Darknet:} The darknet is a part of the internet that is intentionally hidden and not indexed by traditional search engines, often accessible only through specialized software like Tor.
            \item \textbf{Hidden services: } Hidden services, often associated with the Tor network using ".onion" domains, refer to websites and online services that are hosted on servers configured to be accessible only through the Tor network.
            \item \textbf{CSAM:} CSAM, or Child Sexual Abuse Material, refers to explicit media that involves the sexual exploitation or abuse of minors, including images, videos, or other content.
        \end{itemize}
    }%
}
\end{center}

When it comes to CSAM detection, various automatic approaches have been proposed. \citet{automatic_detection} developed a classifier with a true positive rate of 83\% in detecting explicit-like child images and 96.5\% in detecting child faces on a test set of 105 images featuring semi-naked children. \citet{vitorino} utilized convolutional neural networks (CNN) to differentiate regular images from adult and child pornographic content, respectively. \citet{benchmark} created a region-based annotated child pornography dataset (RCPD) in collaboration with the Brazilian Federal Police. They combined face-based child detection with a pornography detector and achieved an accuracy of 79.84\% on the proposed benchmark. Overall, consistently improving CSAM detection algorithms might prompt illegal content creators and distributors to turn to the so-called ``darknet" even more, making it harder for the authorities to assess and prevent CSAM circulation on the web. While the advancement of technology made it easier to moderate and filter abusive and illegal content, it has also provided opportunities for sharing such content with little accountability. CIIVISE states in its interim report that even though France is the fourth largest online host of CSAM in the world, it only employs 1 cyber-crime investigator per 2.2 million people compared to about 1 investigator per 100,000 people in the Netherlands \cite{ciivise2022}.

With its advanced anonymity and privacy features, the Tor network\footnote{\url{https://www.torproject.org/}} has been criticized in the past for facilitating illegal activities in the digital space, including the distribution of CSAM \cite{dw2019}. \citet{gannon_child_2023} find that child abuse sites are 2000 times more prevalent in the darknet, for which Tor provides the main entry point. But they also find that CSAM communities use both the darknet and the clearnet for content sharing: While live streams of child sexual abuse -- predominantly taking place in developing countries -- are mainly hosted in the clearnet, presumably as the risk of law enforcement agencies being aware of live streams is generally perceived to be low, non-live content is predominantly shared via CSAM forums in the darknet. According to \citet{gannon_child_2023}, CSAM-related hidden services usually showcase archaic layouts and do not use high-security technology. Their main protocol to keep the community safe is to share the sites only with like-minded users, typically by invitation from the site administrators or moderators. Some sites require the user to post similar content before they can access the forums. \citet{van2022even} found in a study on a large CSAM forum that while only a fraction of the forum members (0.7 \%) were responsible for 40 \% of the content posted, 9 out of 10 forum members tried to download CSAM at least once.

In this work, we present two major contributions to this field of research: First, to the best of our knowledge, this is the first time that consumption patterns of CSAM are estimated at such a high geographic granularity by correlating it with local-level temporal adult porn consumption patterns. 
Second, we link these fine-granular consumption patterns to both small-area socio-demographic characteristics as well as nearby points of interest and Google Trends\footnote{https://trends.google.com/trends/} queries. While local patterns of both the consumption as well as production of CSAM are relevant for public health professionals and law enforcement agencies alike, we focus on the \emph{consumption} of CSAM for two reasons: First, we assume that uploads of CSAM are mainly done via fixed internet lines/Wifi rather than via the mobile network. Since we only observe mobile network traffic, we consequently expect download traffic to carry stronger signals related to CSAM-related darknet activities. Second, recalling from above, there is a strong empirical link between the consumption of child pornographic content and being involved in sexual violence against children. As \citet{insoll2022risk} points out: 42 \% of survey respondents in his study who have viewed CSAM tried to connect with children online afterwards. Therefore, knowledge about local patterns of CSAM consumption in the darknet may also inform about the prevalence of sexual violence against children in the physical world.

The paper is structured as follows: We describe the data used for this study in Section \ref{section-data}. In Section \ref{section-methodology}, we explain the methodology applied to derive local-level estimates of CSAM consumption and the assumptions used. Commune-level estimates of CSAM consumption for 20 metropolitan regions in France are presented in Section \ref{section-results} alongside their links to POIs, Google Trends and other socio-demographic characteristics. Limitations of the study and words of caution are extensively discussed in Section \ref{section-discussion}.

\section{Data}\label{section-data}
In this study, we aim to analyze local patterns of child pornography consumption for 1341 communes across 20 metropolitan regions in France. The population sizes of the communes in the sample range from 80 in Mont-Saint-Martin, Grenoble to 498,596 in Toulouse averaging at 14,802 across all areas \citep{insee_pop_2019}. The data for Tor usage patterns are derived from geo-referenced, service-level mobile network traffic data measured by the mobile network operator Orange for 20 major cities in France across 77 consecutive days from March 16 to May 31, 2019, provided on a 100$\times$100m spatial grid, also called \textit{tiles} in the following, through the NetMob 2023 data challenge \citep{martínezdurive2023netmob23}. The upload and download data obtained from the mobile network operator is normalized by a random value to conceal the actual traffic of the operator while retaining comparability across web services. Therefore the actual values do \emph{not} have any unit, such as GB, attached to them. It is important to note that mobile network traffic data does not include web traffic generated when connected to fixed-line internet or Wifi. 
The geographic location of a specific user equipment (UE), e.g. a mobile phone, is captured via the base stations of the mobile network the UE is connected with during a given time interval of 15 minutes. The captured web service-specific network traffic is then distributed within the estimated coverage area of the respective base station. For details on the effect of different coverage area estimation approaches we refer to \citet{koebe2020better}. For more details on the data preprocessing performed on the Netmob dataset, we refer to \citet{martínezdurive2023netmob23}.

While data for a variety of web services are provided, we focus on Tor as the main entry point to the darknet. In addition, we consider download traffic from mainly pornographic websites (referred to as `Web Adult' in the following) as a reference for the consumption of pornographic content and download traffic to YouTube as a reference for general mobile video consumption. Both Web Adult and Tor represent multiple web services grouped into a broader category, respectively. However, details on the exact composition of these categories are not available from \citet{martínezdurive2023netmob23}. 

In order to investigate spatial relationships of CSAM consumption and local points of interest (POI), we build on the recently released Overture Maps Foundation (OMF) Places dataset that provides information on about 3 million points of interest for France derived from Meta and Microsoft products such as Bing Maps and Facebook pages \citep{overture}. Using data from OpenStreetMap (OSM) has also been considered, however, OSM provides comparatively little POI information on local businesses. 

Furthermore, we use the reported number of victims of sexual violence as our groundtruth retrieved from the Service Statistique Ministériel de la Sécurité Intérieure (SSMSI) database of the interior ministry of France (Ministère de l'Intérieur et des Outre-Mer) \cite{crimerates}. Socio-demographic information provided by the French National Statistical Office INSEE \citep{insee_pop_2019} and voting outcomes from the 2017 French presidential election are used to control for potential confounders when investigating the link between estimated child pornography consumption and sexual violence.

Lastly, we complement our analysis with information on the relative popularity of search terms from Google Trends. Specifically, we consider the following set of partially community-specific keywords inspired by \citet{owens2022analysis}: \textit{pedoporno}, \textit{porno mineur}, \textit{porno enfant}, \textit{site pedoporno}, \textit{pre-teen hardcore}, \textit{zoo preteen}, \textit{zoo pre-teen}, \textit{pedomom}, \textit{pedodad}, \textit{pthc}, \textit{boylove}, \textit{girllove}, \textit{porno jeune ado}, \textit{video porno ado}, \textit{ado porno}, \textit{porno jeune fille}, \textit{omegle} and \textit{hurtcore}. We extract the relative popularity values of these search terms for each of the 21 regions of France (excluding Corsica, note that Google Trends still uses the regional delineations prior to the 2015 reform) pooled across the years 2017 to 2021 to avoid excessive data sparsity. We map these values to the departments in our sample. While acknowledging that hidden services cannot be found via Google search queries and that the CSAM community actively exchanges ``best practices" to stay anonymous (cf. \citet{gannon_child_2023}), we expect that these keywords may still be able to capture deviances from these practices.

\section{Methodology}\label{section-methodology}

\begin{figure}[ht!]
     \centering
     \caption{\textbf{Composition of CSAM estimates for global and France.}}
    \includegraphics[width=0.45\textwidth]{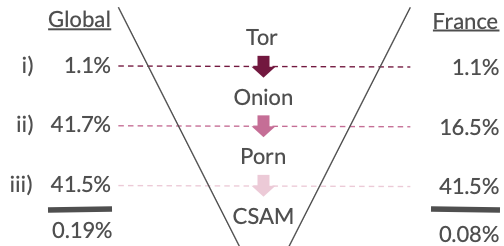}
     \label{fig:funnel}
\end{figure}

In order to narrow down from general Tor usage to child pornography consumption via Tor, we follow a simple, yet effective approach:\\
First, we estimate the global share of CSAM-related Tor traffic by combining three interlinked estimates: i) According to \citet{tortraffic2019}, approx. 
1.1 \% of global Tor traffic went to Onion services during our study period (i.e. March 16 - May 31, 2019). We believe this number to be a conservative estimate for France as \citet{jardine2020potential} report that in `free' countries -- as which France classifies according to Freedom House -- Tor is used more often to access onion services than in the rest of the world. Specifically, they estimate that approx. 7.8 \% of Tor users in free countries use Tor to access onion services vis-à-vis $\sim$6.7 \% on a global level. ii) \citet{jin2023darkbert} collected 5,437,248 of these .onion-pages during the years 2020-22 and observed that the category `Pornography' accounted for approx. 41.7 \% of the collected pages. The authors used the hidden service indexing website Ahmia.fi\footnote{\url{https://ahmia.fi/}} to collect seed addresses for crawling. On the one hand, since Ahmia.fi explicitly blacklists hidden services related to child abuse, we expect that CSAM sites are potentially under-sampled in this dataset (the blacklist contains 40,875 .onion sites as of August 2023). On the other hand, Cloudflare, a major content delivery network and domain name system service provider, allowed Tor browser users from September 2018 onwards to route some of their visits to clearnet websites via one of the ten .onion-addresses of Cloudflare. This could have potentially led to a one-sided increase of onion-traffic that may not have been fully captured by \citet{jin2023darkbert}. However, we cannot observe a substantial increase in the share of onion-traffic to overall Tor traffic between 2017 and the end of 2019 \cite{tortraffic2019}, thus we assume this to have a negligible effect on our approximation. iii) \citet{al2019torank} further disaggregated the category `Pornography' in their DUTA dataset and classified 41.5 \% of .onion websites in this category to be related to child pornography specifically. Consequently, we conclude that approx. 0.19 \% of global Tor download traffic is linked to the consumption of child pornographic content. \\
However, commune-level CSAM consumption in France most likely deviates from global estimates. Thus, in order to locally adapt the global estimate to the 1341 French communes in our study, we use web service-level mobile traffic information from the Netmob dataset. Specifically, we approximate (ii) with the share of Tor traffic related to pornographic content by correlating the observed activity patterns for Web Adult and Tor for each of the 1341 French communes in our sample on an hourly basis across the whole time window of the study using Pearson's $\rho$. The underlying assumption is that the consumption of pornographic content, irrespective of whether adults or children are depicted, follows similar temporal patterns. 
Thus, locations $j$ with a higher temporal correlation are then assumed to have a larger fraction of their Tor traffic related to pornography in general, with $\rho_j = 1$ corresponding to 100\% pornographic content. 
Figure \ref{fig:funnel} illustrates the composition of the estimate for global and France, respectively.

The 16.5 \% for France represents the mean of commune-level correlation coefficients $\rho_j$. To avoid non-sensible \emph{negative} estimates of child pornographic consumption (in the following abbreviated as \textit{cpc}) due to negative correlation coefficients, we replace them with small positive values near zero, denoting it with $\rho\prime_j$. We choose small non-zero replacements to avoid log transformed values going to infinity in later analysis. This affects 14 out of 1341 French communes with negligible effects on the overall distribution. Thus, our commune-level correction factor $c_j$ is defined as $c_j = 0.011 \times 0.415 \times \rho\prime_j$. Table \ref{tab:rho} shows the summary statistics of $\rho_j$ and $c_j$.

\begin{table}[ht!]
\centering
\small
\caption{\textbf{Summary statistics of the potential correction factors.}}
\label{tab:rho}
\begin{tabular}{c|ccccccc}
\textbf{}  & \textbf{mean} & \textbf{std} & \textbf{min} & \textbf{25\%} & \textbf{50\%} & \textbf{75\%} & \textbf{max} \\ \hline
\textbf{$\rho_j$} & 0.1649         & 0.0750        & -0.0493       & 0.1166         & 0.1684         & 0.2139         & 0.4971        \\ \hline
\textbf{$c_j$}   & 0.0008         & 0.0003        & 0.0000        & 0.0005         & 0.0008         & 0.0010         & 0.002       
\end{tabular}
\end{table}

Finally, we define our cpc estimates per 1000 inhabitants for all the $J=1341$ French communes in our sample by

\begin{equation}\label{eq-cpc}
    cpc_j = \frac{c_j \times Tor_j^{DL}}{pop_j} \times 1,000,
\end{equation}

\noindent where $c_j$ denotes the correction factor as described above, $Tor_j^{DL}$ the normalized download traffic related to Tor services and ${pop_j}$ commune-level population counts. An average $c$ of 0.0008 therefore can be interpreted as an estimated 0.08 \% of the observed Tor mobile download traffic in our sample of 20 French metropolitan areas being related to child pornography. 

We consider this to be a conservative estimate of CSAM consumption via Tor for multiple reasons: First, the 41.5 \% refers to the share of pornographic .onion-sites that can be linked to CSAM. However, \citet{owen2015tor} found in 2015, that during the 6-month observation period, sites linked to sexual violence against children accounted for only 2 \% of the hidden services screened in the study, but 82 \% of all requests made via Tor. Second, we assume that image-based content (such as CSAM) largely drives traffic. This assumption is backed by the fact that the top 5 web services in terms of download traffic in the Netmob dataset are predominantly image- or video-based (namely Instagram, Facebook, Netflix, YouTube, Facebook Live) \cite{martínezdurive2023netmob23}. Third, France is the fourth-largest host of online CSAM globally. Assuming a somehow positive relationship between hosting and consuming CSAM, this gives an indication of an overall larger share of CSAM consumption compared to the global average. Lastly and importantly, we assume that pornographic content, especially illegal forms thereof is mainly consumed at home, thus handled via Wifi or a fixed internet line. Thus, this gives indication that the correction factor for France for these internet connection types to be higher.

While directly validating our estimates with information on the actual commune-level consumption of CSAM in France is not possible due to the lack of ground truth data, we indirectly validate our findings by correlating the cpc estimates with an appropriate proxy indicator, in our case commune-level statistics on the number of victims of sexual violence (both adults and minors) per 1000 inhabitants. Recalling the link between child pornography consumption and sexual violence against children indicated by \citet{eke_examining_2011}, \citet{insoll2022risk} and \citet{Hall2007APO} in Section \ref{section-intro} and assuming that a non-negligible fraction of victims of sexual violence are minors, we expect our cpc estimates to show stronger correlations with our proxy than general mobile consumption patterns of e.g. YouTube. However, we stress that this proxy most likely just captures the tip of the iceberg of sexual child abuse: First, the indicator includes rape, attempted rape, and sexual assault including sexual harassment. However, somewhat surprisingly, it does not include sexual abuse, where abuse is distinguished from assault per definition as ``it is carried out without violence, coercion or surprise" \cite{crimerates}. Second, while official numbers report 39,314 victims (minors and adults) of sexual violence in France for the year 2019, \citet{ciivise2022} estimates that 160,000 children alone become victims of sexual violence every year in France, as already noted above. Third, the indicator is only reported for those communes with at least five recorded incidences in three consecutive years in total. This statistical disclosure control measure clearly leads to a non-random selection of communes as large communes are more likely to surpass this threshold. Fourth, local variations in reporting behaviour, especially in small communes with low overall reported numbers, may impact significantly the observed spatial patterns.

Since simple correlations in complex social settings most likely suffer from confounding factors, we build a hierarchical multi-level regression model in order to single out the influence on the number of reported cases of sexual violence that can be uniquely attributed to our cpc estimates, while controlling for a set of potentially relevant other socio-demographic and spatial features. To the best of our knowledge, this is the first attempt to look at large-scale local-level CSAM consumption from a spatial epidemiology perspective. We note that this analysis is exploratory and the presented effects do neither imply a causal relationship nor the directionality of any observed relationship. To underline that both directions of influences are possible, we also present analysis results with our cpc estimates as dependent variable.

In addition, we explore points of interest in 0.1 \% tiles with the highest levels of estimated CSAM consumption. As some of these tiles are located in close proximity to each other, we remove duplicate entries by their unique place identifier. However, we noticed that some places in the OMF Places dataset may still be listed twice, e.g. in two different languages. Thus, duplicate entries may occur, however, we expect these to be negligible. Overture Maps Foundation classifies each POI into categories. We display only those POI categories with $n \geq 3$ in order to limit accidental occurrences on one hand and not to miss out on relevant, but rare categories on the other. To get an estimate for the average download traffic per POI category, we divide the observed download traffic by the number of POIs located for any given tile. In a second step, we average the download traffic across POIs for a given POI category. This leaves us with the average download traffic per POI category. While we acknowledge this to be a crude approximation for the actual traffic generated at a certain POI, we assume that POI categories across the large number of tiles observed are still indicative of existing spatial relationships.

\begin{figure}[ht!]
     \centering
     \caption{\textbf{Association of Google Trends search terms related to child pornography with their principal components for French regions pooled across the years 2017-2021.}}
    \includegraphics[width=0.48\textwidth]{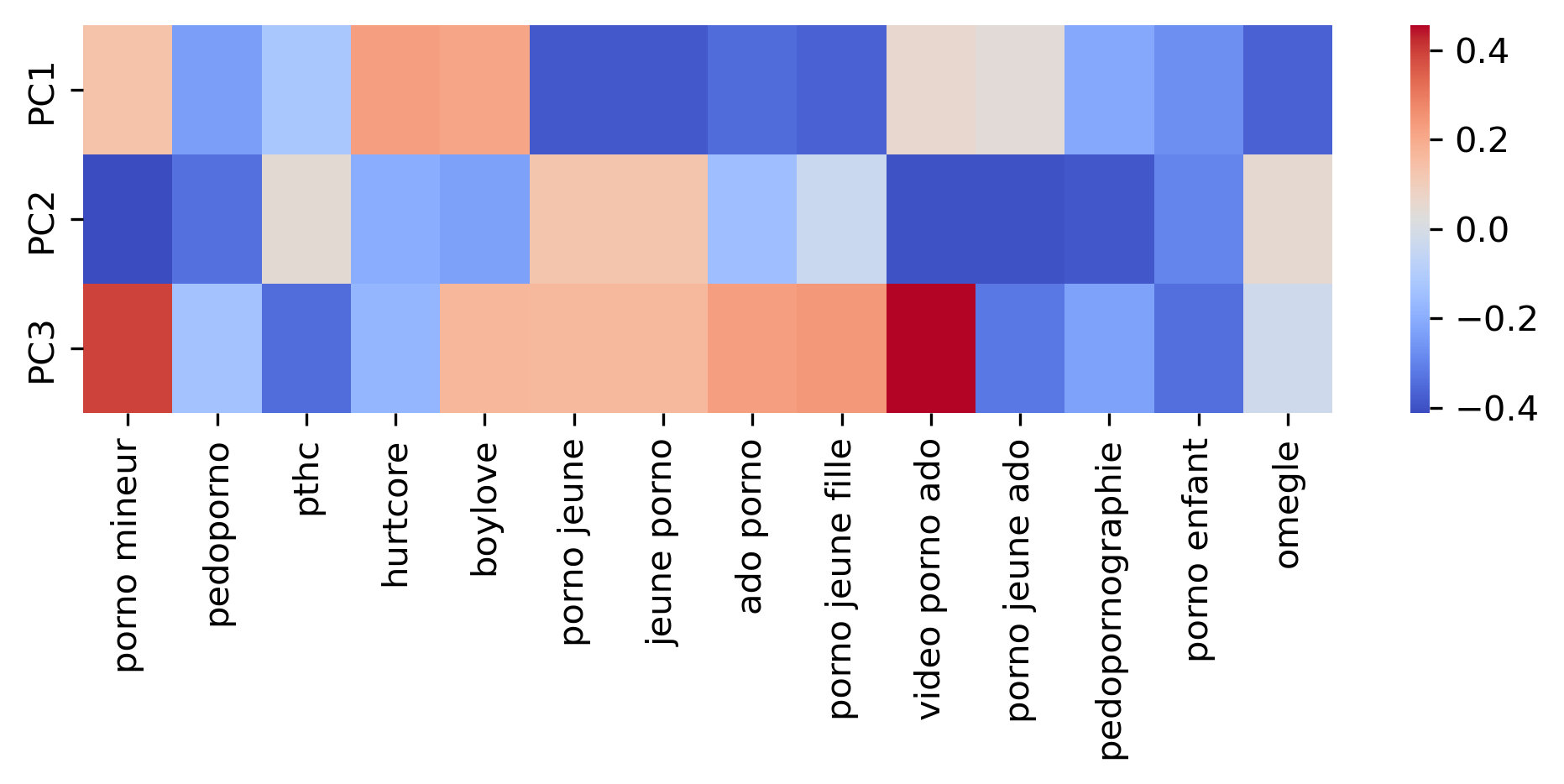}
     \label{fig:pca}
\end{figure}

Of the 18 search terms we extract from Google Trends, we discard seven due to complete sparsity. On the remaining 11 search terms, we perform a principal component analysis with varying number of components. We decided to go for three components by balancing the explained variance and the desired level of abstraction. Figure \ref{fig:pca} shows how the search terms are associated with each of the three components.

Of the three components, we just consider the first (\textit{PC1}) and the third (\textit{PC3}) in further analysis as they appear to capture sexual preferences towards children more succinctly.

\section{Results}\label{section-results}
\begin{figure}[ht]
     \centering
     \caption{\textbf{Normalized download traffic aggregated to day-of-week by hour-of-day, by web service.}}
     \begin{subfigure}[c]{1\linewidth}
         \centering
         \caption{YouTube}
         \includegraphics[width=1\textwidth]{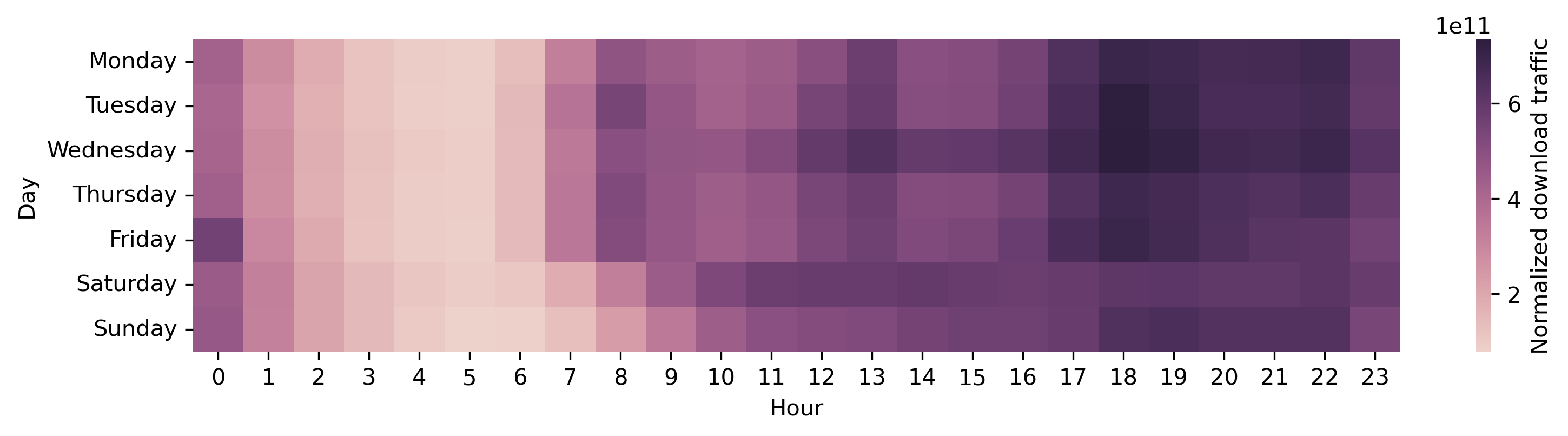}
         \label{fig:heatmap_yt}
     \end{subfigure}
     \hfill
     \begin{subfigure}[c]{1\linewidth}
         \centering
         \caption{Web Adult}
         \includegraphics[width=1\textwidth]{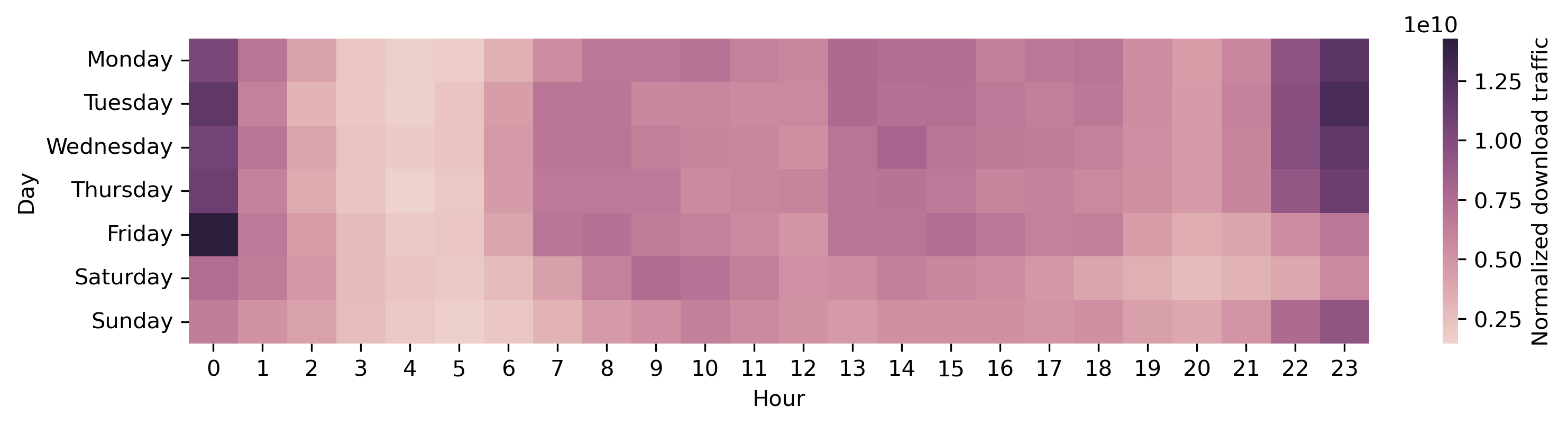}
         \label{fig:heatmap_wa}
     \end{subfigure}
     \hfill
     \begin{subfigure}[c]{1\linewidth}
         \centering
         \caption{Tor}
             \includegraphics[width=1\textwidth]{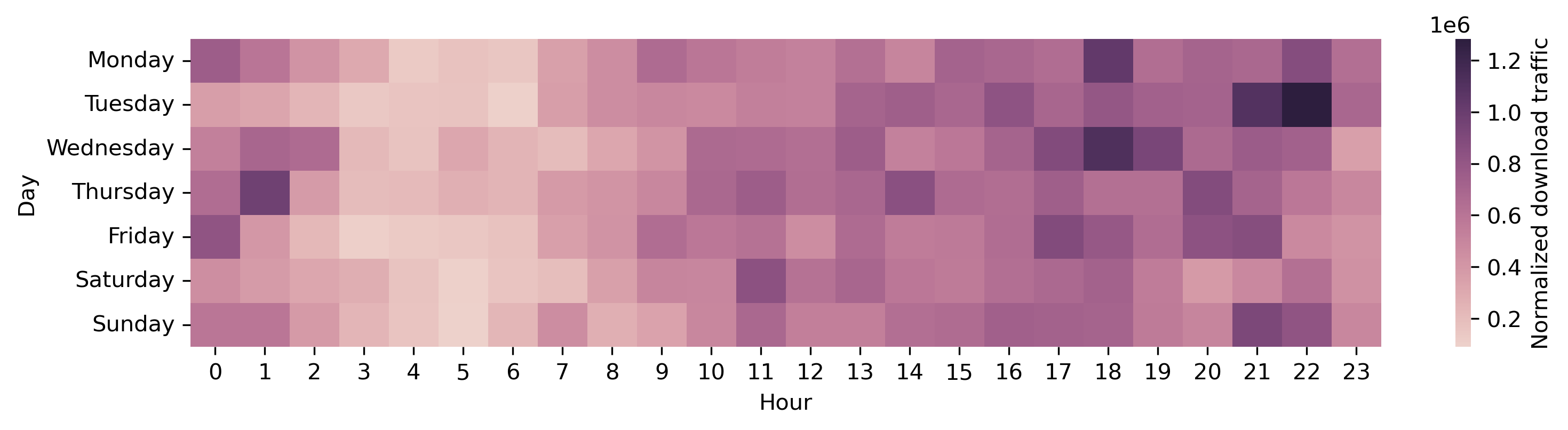}
         \label{fig:heatmap_tor}
     \end{subfigure}
     \hfill
     \begin{subfigure}[c]{1\linewidth}
         \centering
         \caption{CPC estimates - Top 10 communes}
         \includegraphics[width=1\textwidth]{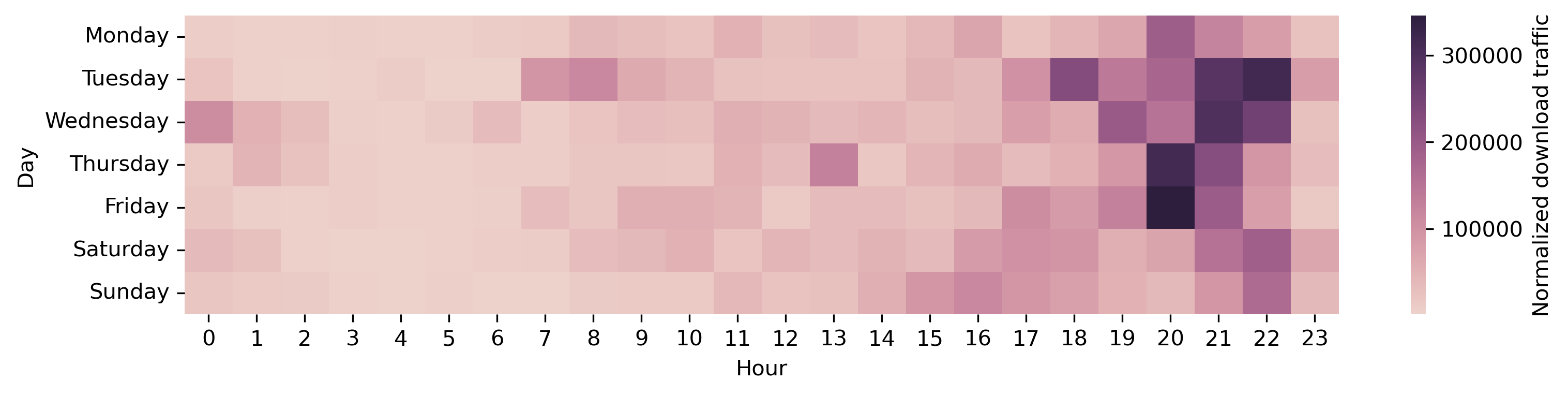}
         \label{fig:heatmap_tor_10}
     \end{subfigure}
     \label{fig:heatmap}
\end{figure}

Estimated CSAM consumption per 1000 inhabitants ranges from 0 in 14 communes to 157,077 in Mondouzil, Toulouse averaging 3,703 across all areas between March 16 to May 31, 2019. As noted above, there is no actual unit attached to the traffic volume as it is normalized by the mobile phone operator. For comparison, YouTube download traffic per 1,000 inhabitants averages 3,743,939,828 across all areas during the same time window, thus more than a million times the average Tor download traffic estimated to be related to CSAM. Commune-level results displayed in Figure \ref{fig:cpc_main} in the Appendix. 
While more fine-granular estimates, e.g. on the tile-level (100m  or census district (IRIS)-level, are technically possible, the share of census population estimates close to zero grows dramatically for small areas, thus rendering lower-level estimates per 1,000 inhabitants increasingly volatile. Therefore, we opt to present commune-level estimates in this study. However, as we observe \emph{mobile} internet traffic only, the locations of (i) the traffic generation and (ii) the place of residence of the user do not necessarily coincide. 
Although we account for varying population sizes across communes, we observe that tile-level activity patterns are not necessarily propagated and visible on the commune-level. In other words, highly active tiles do not lead to highly active communes in terms of Tor download traffic, especially if these communes are large. This hints at spatially highly concentrated traffic generation. This argument is also supported when looking at Figure \ref{fig:heatmap}, which shows the normalized download traffic for YouTube, web adult content, and Tor services summarized by weekday and hour across all cities in the sample.

As one might expect, all of the services analyzed show major peak traffic in the evening hours outside of regular business hours, thus hinting at the private entertainment purpose of these services. Download traffic from YouTube and adult content vary smoothly across the hours of the day with additional subtle peaks around 8am and 1pm during weekdays. CPC-related traffic in the 10 communes with the highest CSAM consumption estimates shows a stronger concentration of download activity in the evening hours compared to overall Tor download traffic. However, Tor-based traffic appears more coarse-grained in general. A potential explanation for the pixelated appearance of Tor-based download traffic is that Tor services saw approx.~2.5 million daily visitors globally in 2022 \citep{tortraffic2022}, while the general internet is used by approx.~five billion users per day in 2022 \citep{itu2022}. The Tor project estimates 100,537 mean daily Tor users for France during the time window of our study. Thus, it is likely that local-level Tor mobile download traffic via one mobile network operator is driven by a comparatively small subscriber base in our sample, so individual uses have a larger effect on the aggregate.

\subsection{Validating estimates against official statistics on sexual violence}\label{section-validation}

While direct validation of our methodology is hardly possible due to the lack of statistical data on child porn consumption habits, we indirectly validate our findings by correlating our cpc estimates with commune-level statistics on the number of victims of sexual violence per 1000 inhabitants as described in Section \ref{section-methodology}. Looking back at \cite{eke_examining_2011} and \cite{Hall2007APO} in Section \ref{section-intro} that link child pornography consumption and sexual violence, we expect our cpc estimate to indicate a positive association with the reported number of victims of sexual violence than general mobile consumption patterns. Table \ref{tab:corr_violence} shows the correlations of the number of victims of sexual violence with download traffic of YouTube, Web Adult, Tor, and cpc estimates, respectively, and whether these correlations are significantly different from zero.

\begin{table}[ht]
\centering
\small
\caption{\textbf{Correlation of the reported number of victims of sexual violence with log download traffic, per 1000 inhabitants at the commune-level and by web service.} For $n = 731$ communes for which official numbers are available.}
\begin{tabular}{c|c|c}
Indicator & Spearman's $\rho$ & p-value \\
\hline
log\_YouTube\_per\_1000 & 0.178 & 1.33e-06 \\
log\_Web\_Adult\_per\_1000 & 0.161 & 1.18e-05\\
log\_Tor\_per\_1000 & 0.237 & 7.94e-11 \\
log\_cpc\_per\_1000 & 0.284 & 4.99e-15\\
\hline
\end{tabular}
\label{tab:corr_violence}
\end{table}

In addition, we perform paired-samples tests for dependent correlation coefficients to check whether the correlation coefficient of our cpc estimates with the reported number of victims of sexual violence differs significantly from the other three web services. We see that the cpc estimates correlate significantly stronger with the number of victims of sexual violence (per 1,000 inhabitants) than the other three web services (all three p-values < 1e-08). However, relying on correlations to investigate complex social phenomena is prone to confounding influences. Consequently, in further analysis, we link our commune-level cpc estimates to socio-demographic characteristics and other expectedly relevant spatial factors. To do so, we collect demographic data at the levels of communes, intercommunalities, and departments in France from the French statistical office INSEE including data on voting behaviour during the 2017 French presidential election, and combine them with the number of certain POIs per 1000 inhabitants and sets of Google Trends search terms related to (child) pornography. We chose the POIs based on the argument by \cite{sauve2021violences} that sexual violence against children mostly happens in places where a lot children are, e.g. at home, in schools or in sports clubs. Although child abuse and CSAM consumption may not happen at the same location, it is feasible to assume that offenders are in most cases not strangers to those places and likely live nearby, i.e. in the same commune. As \cite{ciivise2022} states: In France, 8 out of 10 victims of child sexual abuse are victims of incest, in most cases committed by the older brother or father. Although both directions of the effect between our cpc estimates and the reported number of victims of sexual violence are plausible and supported by academic literature (cf. Section \ref{section-intro}), we cannot determine the directionality of the relationship in our study design. Thus, we provide results for both directions by fitting one indicator on the other while controlling for a set of potential confounders using an ordinary least squares model with heteroscedasticity-robust standards errors. The results are presented in Table \ref{tab:ols}.

\begin{table}[!htbp] \centering 
\small
  \caption{\textbf{Regression results for our cpc estimates and the number of victims of sexual violence averaged across 2017-21, by commune.}} 
  \label{tab:ols} 
\begin{tabular}{@{\extracolsep{5pt}}lcc} 
\\[-1.8ex]\hline 
\hline \\[-1.8ex] 
 & \multicolumn{2}{c}{\textit{Dependent variable (per 1000):}} \\ 
\cline{2-3} 
\\[-1.8ex] & log\_cpc & Sexual\_violence\\ 
\hline \\[-1.8ex] 
 log\_YouTube\_per\_1000 & 0.35$^{***}$ & 0.04 \\ 
  & (0.11) & (0.09) \\ 
 log\_Web\_Adult\_per\_1000 & $-$0.25$^{**}$ & $-$0.09 \\ 
  & (0.10) & (0.10) \\ 
 log\_Tor\_per\_1000 & 1.13$^{***}$ & $-$0.10 \\ 
  & (0.04) & (0.08) \\ 
 log\_pop\_density & 0.14$^{***}$ & 0.02 \\ 
  & (0.03) & (0.08) \\ 
 Share\_of\_singles & 0.80 & 3.82$^{**}$ \\ 
  & (0.66) & (1.75) \\ 
 Poverty\_rate & $-$5.21$^{***}$ & 2.02 \\ 
  & (1.17) & (1.58) \\ 
 Employment\_rate & $-$8.40$^{***}$ & 0.92 \\ 
  & (2.24) & (1.39) \\ 
 Electoral\_turnout\_2017 & $-$2.27$^{**}$ & $-$1.12$^{***}$ \\ 
  & (0.96) & (0.40) \\ 
 Share\_Le\_Pen & 0.59 & 0.73 \\ 
  & (1.02) & (0.57) \\ 
 Share\_Macron & 2.72$^{*}$ & 0.05 \\ 
  & (1.54) & (0.64) \\ 
 POI\_adult\_entertainment & $-$0.003 & 0.02 \\ 
  & (0.01) & (0.03) \\ 
 POI\_sports\_teams & $-$0.001 & 0.03 \\ 
  & (0.01) & (0.04) \\ 
 POI\_church & $-$0.01 & $-$0.02 \\ 
  & (0.01) & (0.02) \\ 
 POI\_mosque & $-$0.03$^{**}$ & 0.003 \\ 
  & (0.01) & (0.02) \\ 
 POI\_religious\_org & $-$0.06$^{*}$ & 0.02 \\ 
  & (0.04) & (0.02) \\ 
 POI\_school & $-$0.02 & 0.07$^{*}$ \\ 
  & (0.02) & (0.04) \\ 
 PC1 & $-$0.03$^{*}$ & 0.01 \\ 
  & (0.02) & (0.01) \\ 
 PC3 & $-$0.01 & $-$0.04$^{**}$ \\ 
  & (0.03) & (0.02) \\ 
 Sexual\_violence\_per\_1000 & 0.14$^{**}$ &  \\ 
  & (0.07) &  \\ 
 log\_cpc\_per\_1000 &  & 0.10$^{*}$ \\ 
  &  & (0.06) \\ 
 Constant & $-$11.04$^{***}$ & 1.26$^{**}$ \\ 
  & (1.00) & (0.52) \\ 
\hline \\[-1.8ex] 
Observations & 630 & 630 \\ 
R$^{2}$ & 0.96 & 0.41 \\ 
Adjusted R$^{2}$ & 0.96 & 0.39 \\ 
\hline 
\hline \\[-1.8ex] 
\textit{Note: Heteroscedasticity-robust SE}  & \multicolumn{2}{r}{$^{*}$p$<$0.1; $^{**}$p$<$0.05; $^{***}$p$<$0.01} \\ 
\end{tabular} 
\end{table} 

We observe that both the cpc estimates as well as the sexual violence indicator have a small, but positive and statistically significant impact on the respective outcome. Also, we see that the overall explained variance measured in (adjusted) R\textsuperscript{2} is higher for cpc estimates than for the sexual violence indicator. This is expected as we control for download traffic of related web services. Interestingly, the effect of adult porn consumption (\textit{log\_Web\_Adult\_per\_1000}) is negative, which hints at a subtle substitution effect: adult porn consumption in the clearnet is to some extent replaced by CSAM consumption in the darknet. Furthermore, we notice little consistency with regard to the direction, significance, and size of the observed effects of the control variables across the two regression setups. Together, this hints at the fact that our cpc estimates and the sexual violence capture two distinct behaviours. While this could either support or undermine the validity of our estimate -- we are able to single out the signal related to CSAM from the noisy sexual violence indicator vis-à-vis we measure some completely different Tor usage behaviour -- both the positive and significant association of the two indicators as noted above and the fact that we control for overall Tor download traffic supports the validity of our estimates. To investigate this further, we repeat the analysis for various specifications (see Appendix). We use the reported cases of drug abuse per 1000 inhabitants as a proxy for another presumably popular use of the darknet -- ordering drugs. As Table \ref{tab:robust-drug} in the Appendix shows, the drug abuse rate does not inform our cpc estimates, giving further indication that we capture (child) porn-related consumption as we do not capture marketplace-related uses of Tor.

Further, we observe that the sexual violence indicator is zero for approx. half of the communes in our sample. Zero inflation may bias our parameter estimates as it hints at unmodelled factors causing the zeros in the first place. In Table \ref{tab:robust-nozero} in the Appendix, we therefore exclude the communes with no reported cases to check for the impact of a zero-inflated setting. Overall, the significance of the observed effects is reduced which can be to some extent explained by the reduction in sample size, but significant effects do not show a change in sign or size.
Lastly, it needs to be pointed out that the level of variation attached to these findings are most likely vastly underestimated, since the uncertainty involved in both the approximation of the correction factor as well as the underreporting of sexual abuse/violence cases is not accounted for, just to name a few. Also, we would like to stress, that this analysis does not, in any way, indicate that people of certain demographics participate in child abuse. Rather, our results should be interpreted as a first step into little-charted territory, namely looking at sexual child abuse via CSAM consumption from a spatial epidemiological perspective.



\subsection{Investigating spatial relationships of child sexual abuse materials}\label{section-poi}

\begin{figure}[ht]
     \centering
     \caption{\textbf{Commune with the highest cpc estimate.} The line plot describes the hourly cpc estimates for each tile within the respective commune. The heatmap shows the cpc estimates by weekday and hour.}
     \begin{subfigure}[c]{1\linewidth}
         \centering
         \includegraphics[width=1\textwidth]{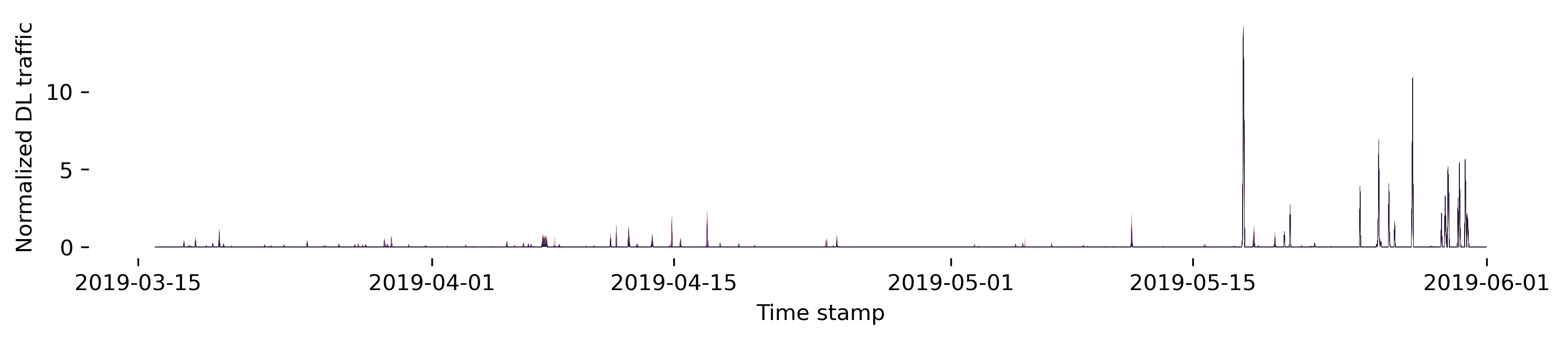}
         \label{fig:timeline_tor_1}
     \end{subfigure}
     \hfill
     \begin{subfigure}[c]{1\linewidth}
         \centering
         \includegraphics[width=1\textwidth]{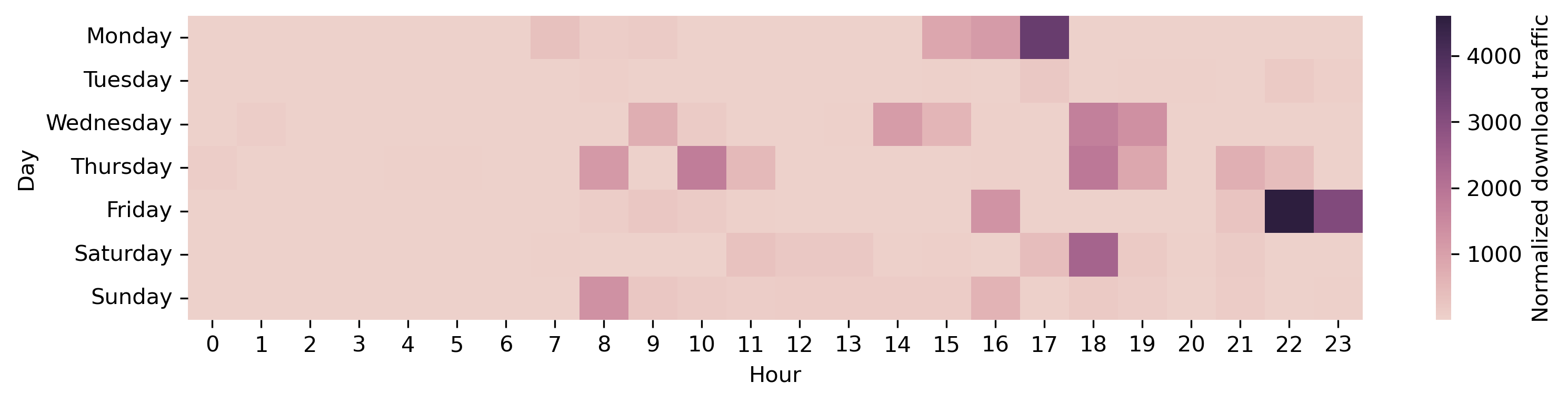}
         \label{fig:heatmap_tor_1}
     \end{subfigure}
     \label{fig:tor_highest}
\end{figure}

We further investigate the spatial relationship of estimated CSAM consumption with the local environment. In Figure \ref{fig:tor_highest}, we present the cpc estimates of the 
commune for which we estimate the highest CSAM consumption per 1,000 inhabitants in our sample of 1341 communes. Tile-specific Tor download traffic is multiplied by the respective commune-level correction factor. The correction factor does not vary over time, but has been calculated for the whole time window of our study.

By looking at the timeline of Tor download traffic 
in this commune, Tor services appear to be used rather irregularly, as already mentioned before. Thus, we not only see a spatially, but also temporarily highly concentrated Tor usage. This would be in line with some common CSAM practices as described by \cite{gannon_child_2023}, where CSAM is usually not streamed on-demand, but downloaded and consumed offline. While this may explain the ``front-loaded" cpc download activity apparent in Figure \ref{fig:heatmap_tor_10} when compared to adult porn download activity in Figure \ref{fig:heatmap_wa} and therefore validates our main assumption that porn consumption follows the same temporal pattern, regardless whether adults or children are depicted, it lays open a caveat in it: porn consumption and consumption-related download traffic do not necessarily occur simultaneously, especially in the CSAM community. Based on the visual inspection of Figures \ref{fig:heatmap_tor_10} and \ref{fig:heatmap_wa}, we determine the potential time lag between activity and assumed consumption to be around two hours. Consequently, we lag Tor traffic by two hours and re-run both the calculation of the correction factor and the subsequent regression analysis. The lagged Tor traffic improves our pairwise correlation with the sexual violence indicator as reported in Table \ref{tab:corr_violence} from 0.28 to 0.34 as well as our regression analysis as presented in Table \ref{tab:robust-lag} of the Appendix. Even though the patterns observed via the day-of-week by hour-of-day heatmap does not indicate a generalizable usage pattern, one can clearly see that it does not align with regular business hours and therefore indicate private use. This argument is supported by the fact that the most active tiles within the commune displayed here are located in residential or rural neighborhoods as visual inspection of the respective tile locations on Google Earth shows. 

By looking not only at the top 10 communes with the highest estimated CSAM consumption, but at the 0.1 \% of all tiles with the highest download traffic (n = 5,259) for the three different web services in our study, we observe distinct sets of adjacent points of interest (POIs) as shown in Table \ref{tab:poi}.

\begin{table}[ht]
\small
\centering
\caption{\textbf{Points of interest in the 0.1 \% of tiles with the highest cpc traffic, by web service.} Number of POIs in parentheses.}
\label{tab:poi}
\begin{tabular}{cccc}
\multicolumn{1}{c|}{\textbf{Rank}} & \multicolumn{1}{c|}{\textbf{CPC}}                       & \multicolumn{1}{c|}{\textbf{Web Adult}} & \textbf{YouTube} \\ \hline
\multicolumn{1}{c|}{1} & \multicolumn{1}{c|}{embassy (37)} 
                        & \multicolumn{1}{c|}{jazz\_and\_blues (3)}
                        & tennis\_court (3) \\
\multicolumn{1}{c|}{2} & \multicolumn{1}{c|}{lake (4)} 
                        & \multicolumn{1}{c|}{jail\_and\_prison (3)} 
                        & attractions\_and\_activities (17) \\
\multicolumn{1}{c|}{3} & \multicolumn{1}{c|}{contract\_law (5)}
                        & \multicolumn{1}{c|}{b2b\_services (3)} 
                        & aquarium (3) \\
\multicolumn{1}{c|}{4} & \multicolumn{1}{c|}{swimming\_pool (9)} 
                        & \multicolumn{1}{c|}{roofing (3)} 
                        & boat\_tours (3) \\
\multicolumn{1}{c|}{5} & \multicolumn{1}{c|}{eye\_care\_clinic (8)} 
                        & \multicolumn{1}{c|}{agricultural\_cooperatives (3)} 
                        & mining (3) \\ \hline
\end{tabular}
\end{table}

Although one could think of plausible explanations for some of the POIs in Table \ref{tab:poi} (e.g. concerning the use of web services related to adult pornography around prisons or the use of YouTube at tourist attractions), drawing more general spatial relationships from Table \ref{tab:poi} appears challenging, especially for our CPC estimates. For example, it is unclear whether Tor plays an important role in fulfilling diplomatic duties or whether these high levels of Tor mobile download traffic are simply a geographic coincidence. An argument against the latter is that this coincidence is not limited to one larger diplomatic area, but occurs across several cities in France. A detailed look at the POI locations for our CPC estimates reveals that many of the POIs across the mentioned categories are located around Porte de Passy, which surrounding area represents the largest Tor download traffic hotspot in our sample of 20 urban areas in France. However, most of the CPC-related traffic in the corresponding commune is generated throughout the study period at the end or outside of regular office hours. 
Noticeable is that many of the identified POIs are located in densely populated areas. One explanation for that is that we look at total traffic on the tile-level as tile-level population statistics are on one hand not readily available and on the other hand potentially misleading, especially in tourist areas. Interestingly, a closer look at the actual POI locations also reveals generally fewer POI locations in the OMF Places dataset vis-à-vis Google Maps.

Importantly, it needs to be stressed here that just because traffic is generated in close proximity to these places, it does not mean that this traffic is generated by the inhabitants, owners, or employees themselves, but by any subscriber near the location. Related to the well-known concept of ecological fallacy, area-level correlations do necessarily not imply individual or POI-level causal relationships. As an example, while prostitution occurs mainly in poorer neighborhoods, the clients are not necessarily the poor locals.  

\section{Discussion}\label{section-discussion}



In this study, we shed light on a topic usually hidden in the dark from a novel angle: We looked at spatial patterns in the consumption of child sexual abuse material using mobile network data for 1341 small areas across 20 metropolitan areas in France for 77 consecutive days in 2019. To the best of our knowledge, this is the first time that spatial CSAM consumption patterns have been mapped at such a high geographical detail. Validated against the reported numbers of victims of sexual violence at the commune-level, we further explored geographic links to both local socio-demographic characteristics as well as to nearby points of interest and Google search queries. These insights may contribute to a better understanding of the whereabouts of CSAM consumption and thus inform targeting public awareness campaigns such as the one launched in September 2023 by the French government \citep{lemonde2022}. However, it is important to address the limitations inherent to this study: First, the study analyzes mobile network traffic from one major mobile network provider only; hence, it misses out on web traffic generated both via Wifi or fixed internet connections or via other mobile network operators. Structural differences between mobile-only and overall traffic, especially when it comes down to the consumption of (child) pornography, need to be expected but cannot be further quantified in this study. Second, our estimates build on assumptions as laid out in Section \ref{section-methodology}, since detailed information concerning the specific origin of the observed Tor download traffic is not available. While we try to support the assumptions with evidence, they may not hold to a full extent, especially on local levels, as the sample size of actual Tor users generating the observed traffic might be very small. Third, linking consumption patterns with local phenomena such as socio-demographic characteristics or points of interest is subject to additional uncertainty as the mobile traffic is assumed to be generated partly out-of-home, i.e. not exclusively by the inhabitants of that area, but potentially by any visitor. Therefore, relationships observed on the area-level may not hold on the individual-level. Fourth, the sourcing of the POI information is not described in detail by the data provider and, therefore, may be prone to certain selection biases. Especially as residential homes are usually not counted as a point of interest, information on these might be underrepresented or captured indirectly by POIs prevalent in residential areas such as schools. Fifth, our groundtruth indicator, i.e. the reported number of victims of sexual violence, is imperfect in many ways as laid throughout the study, but -- to the best of our knowledge -- the most suitable proxy for child sexual abuse in France on local levels. Consequently, our CSAM-related consumption estimates need to be considered with caution, especially at the local level.

As described in the Netmob dataset description, data collection, processing and aggregation took place in compliance with GDPR under the supervision of the Data Protection Officer of the mobile network operator Orange \citep{martínezdurive2023netmob23}. Individual-level traffic has been aggregated to 15-minute intervals and spatially distributed across a network coverage grid. Furthermore, the study authors refrain from any detailed depiction of small areas, e.g. presenting geographic coordinates for single tiles that could put people or businesses at risk of being accused of wrongdoing. In addition, we tried to add flags of caution throughout the study to avoid that individual figures or paragraphs can be misinterpreted when taken out of context.

Going ahead, we see multiple ways how this research can be extended: First, the regression could benefit from additional indicators that capture attitudes, behaviours, and opinions in a more nuanced way. This is of particular importance for deriving policy implications from our work. 
Second, we have not found any major external shock such as take-downs of large CSAM forums in the darknet during the time window of analysis. Re-running the analysis around such an event may provide further insights into the agility and resilience of the community to external interventions. Third, in a related manner, temporal information on forum activities may help to link specific forum activities (e.g. release of a new curated CSAM collection) with traffic patterns. Fourth, extending the analysis to fixed internet connections may allow to capture the full extent of CSAM consumption online and help to quantify the bias induced by observing mobile traffic only. This would also allow to investigate the \emph{supply} side of the CSAM market more rigorously, namely the upload traffic. Lastly, we hope that the release of the Netmob dataset will strike a precedent for other mobile network operators and internet service providers to provide web service-level network traffic information to researchers in an ethical manner. While the internet has fundamentally transformed the way we behave and communicate, it is still little known how it is actually used in everyday life. Consequently, more such data releases would facilitate research not only on the darknet, but across a wide range of disciplines.


In conclusion, we believe that our study sheds light on the consumption of CSAM from a novel angle using so far little-tapped data source -- large-scale web service mobile traffic. In that way, we hope that our study can help in better understanding the spatial relationship between CSAM consumption and child sexual abuse and ultimately help to move forward on target 16.2 of the Sustainable Development Goals: "End abuse, exploitation, trafficking and all forms of violence and torture against children".



\bibliography{references.bib}

\section*{Acknowledgements}
We are greatly thankful for the valuable feedback of Roger Dingledine on the initial version of the paper.

\appendix

\section{Appendix}

\newpage

\begin{figure}[htbp]
     \centering
     \caption{\textbf{Commune-level estimates of child pornography consumption per 1000 inhabitants, by major French metropolitan area. See the Discussion section for limitations of the underlying analysis.}}
     \begin{subfigure}[c]{0.18\linewidth}
         \centering
         \includegraphics[width=1\textwidth]{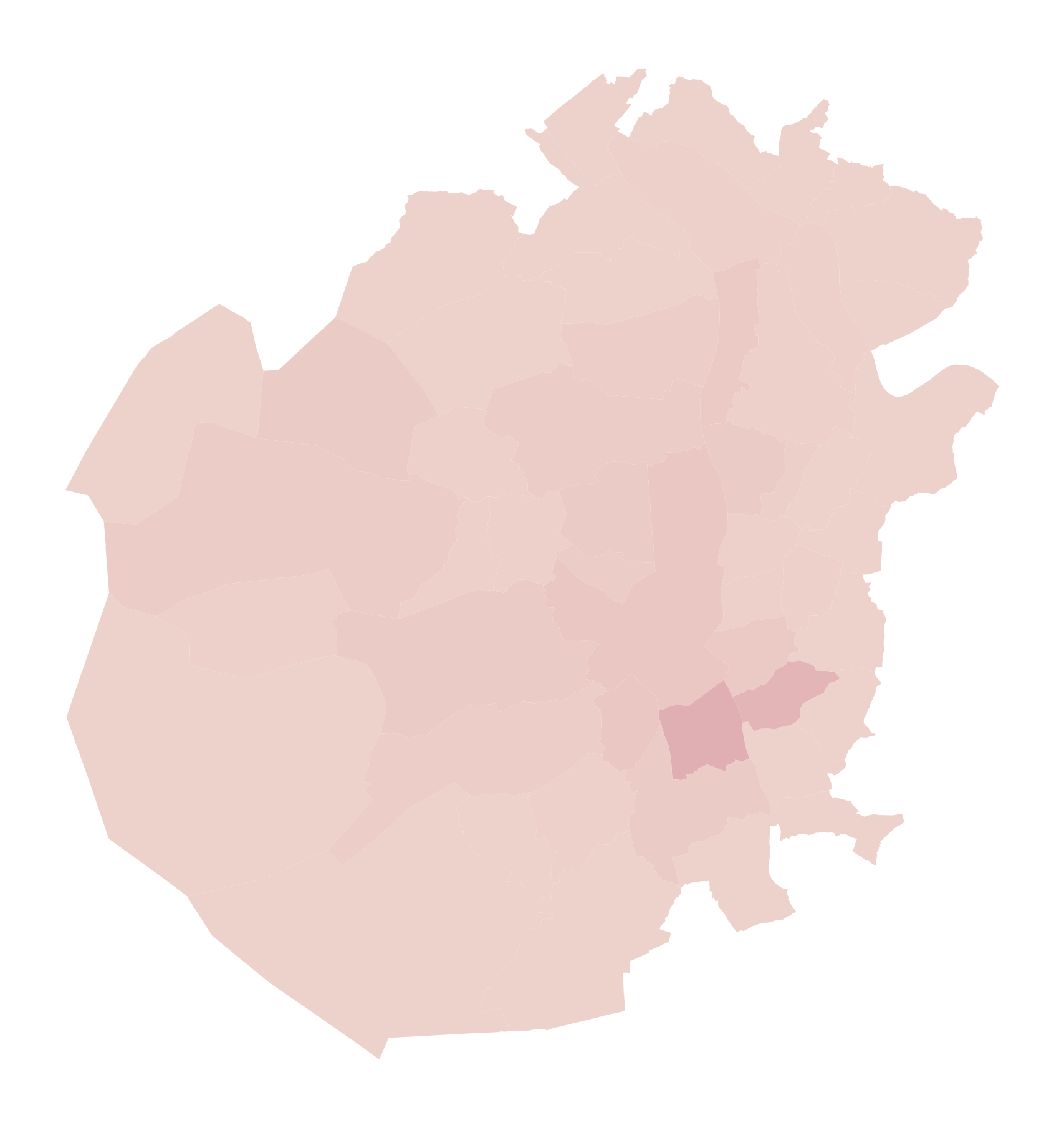}
         \caption{Bordeaux}
         \label{fig:cpc_bordeaux}
     \end{subfigure}
     \begin{subfigure}[c]{0.18\linewidth}
         \centering
         \includegraphics[width=1\textwidth]{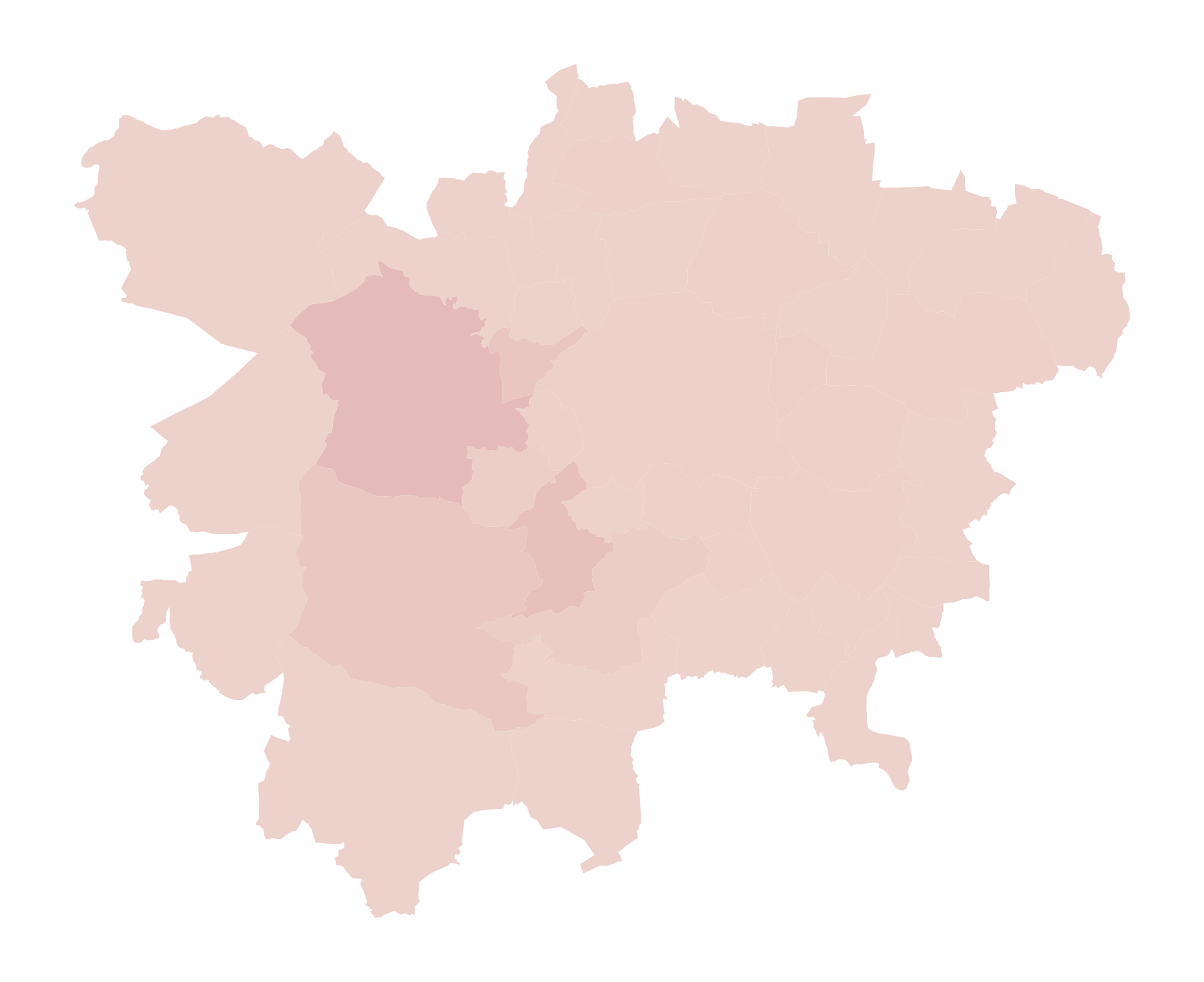}
         \caption{Clermont-Ferrand}
         \label{fig:cpc_Clermont-Ferrand}
     \end{subfigure}
     \begin{subfigure}[c]{0.18\linewidth}
         \centering
         \includegraphics[width=1\textwidth]{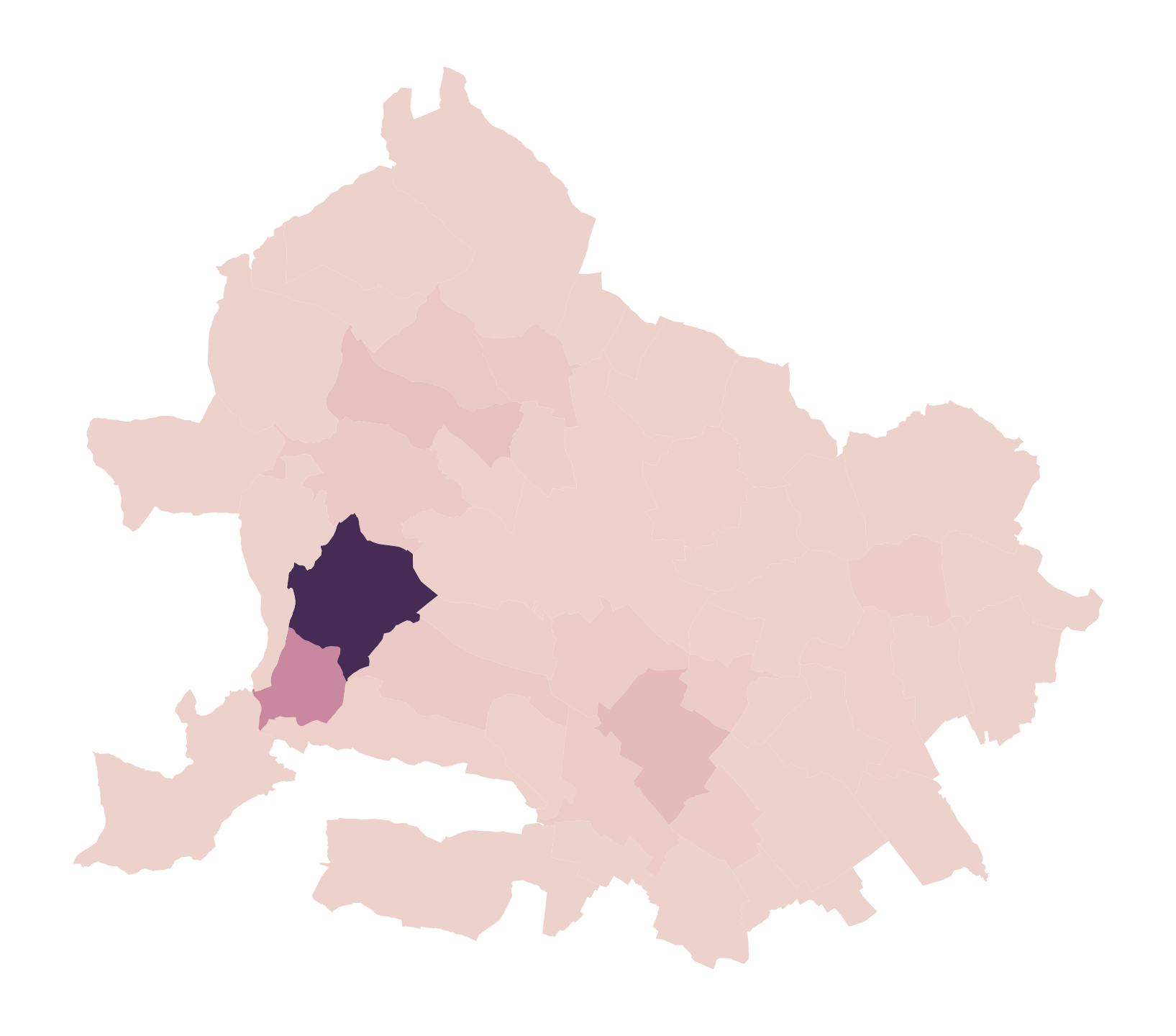}
         \caption{Dijon}
         \label{fig:cpc_Dijon}
     \end{subfigure}
     \begin{subfigure}[c]{0.18\linewidth}
         \centering
         \includegraphics[width=1\textwidth]{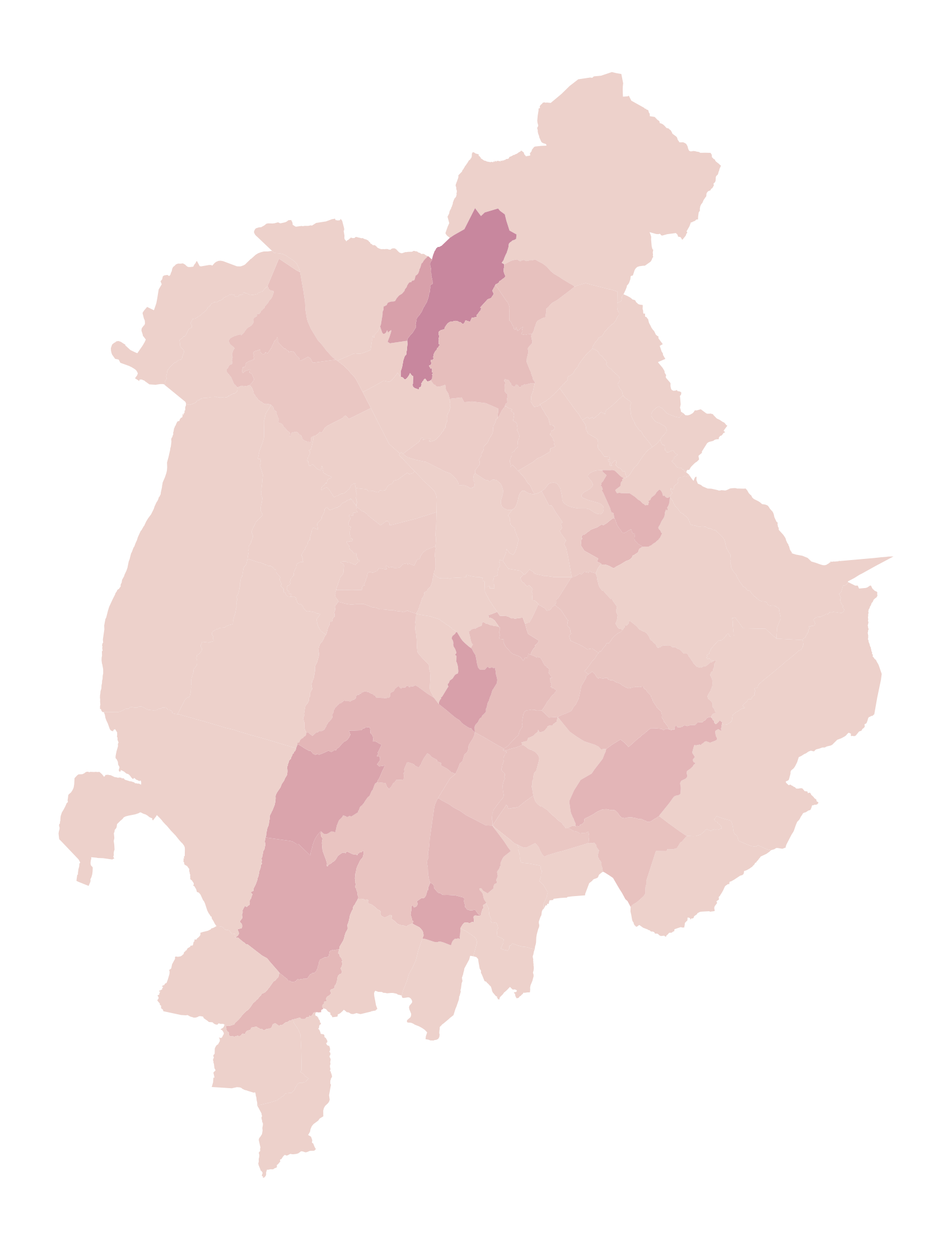}
         \caption{Grenoble}
         \label{fig:cpc_Grenoble}
     \end{subfigure}
     \hfill
     \begin{subfigure}[c]{0.18\linewidth}
         \centering
         \includegraphics[width=1\textwidth]{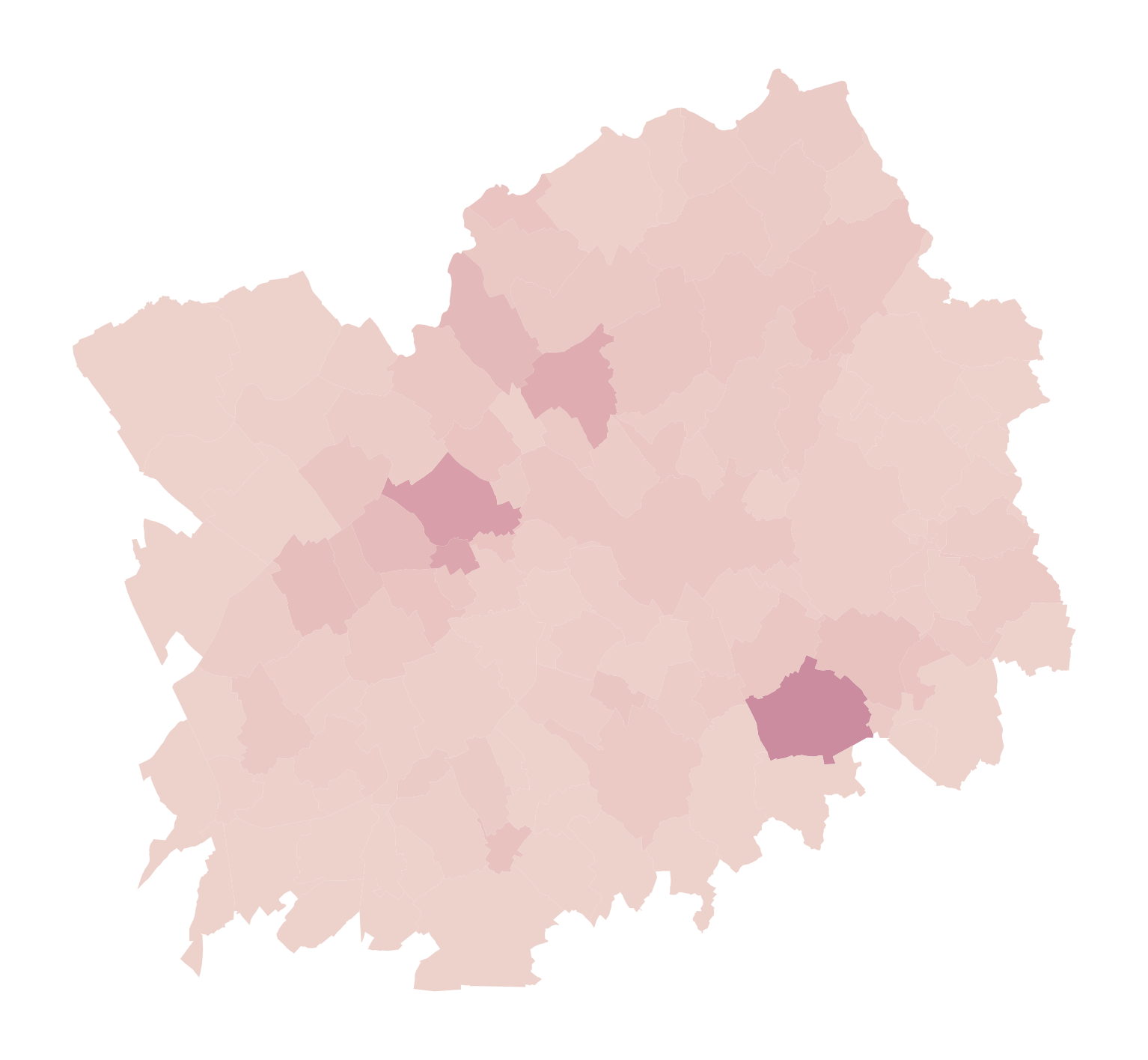}
         \caption{Lille}
         \label{fig:cpc_lille}
     \end{subfigure}
     \begin{subfigure}[c]{0.18\linewidth}
         \centering
         \includegraphics[width=1\textwidth]{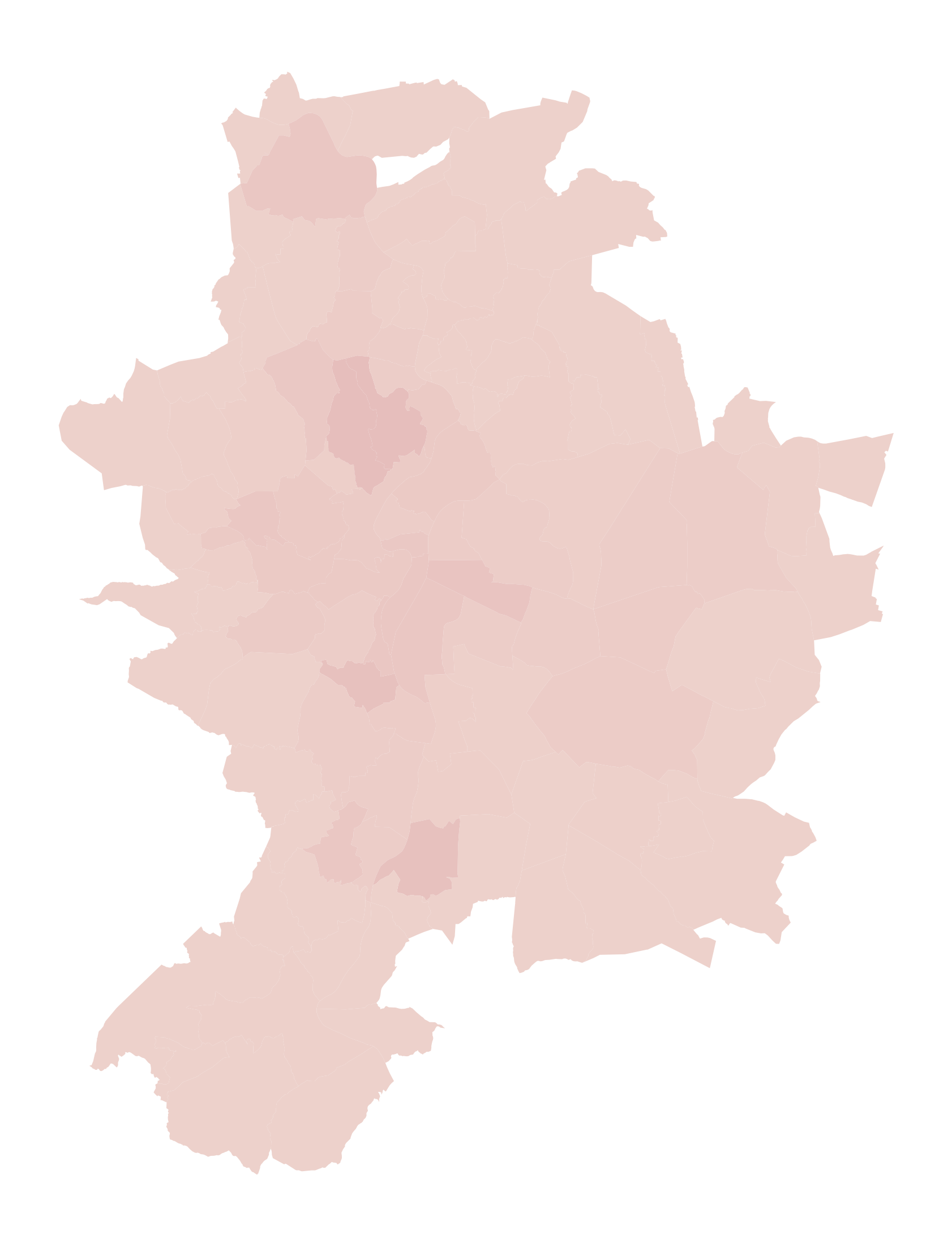}
         \caption{Lyon}
         \label{fig:cpc_Lyon}
     \end{subfigure}
     \begin{subfigure}[c]{0.18\linewidth}
         \centering
         \includegraphics[width=1\textwidth]{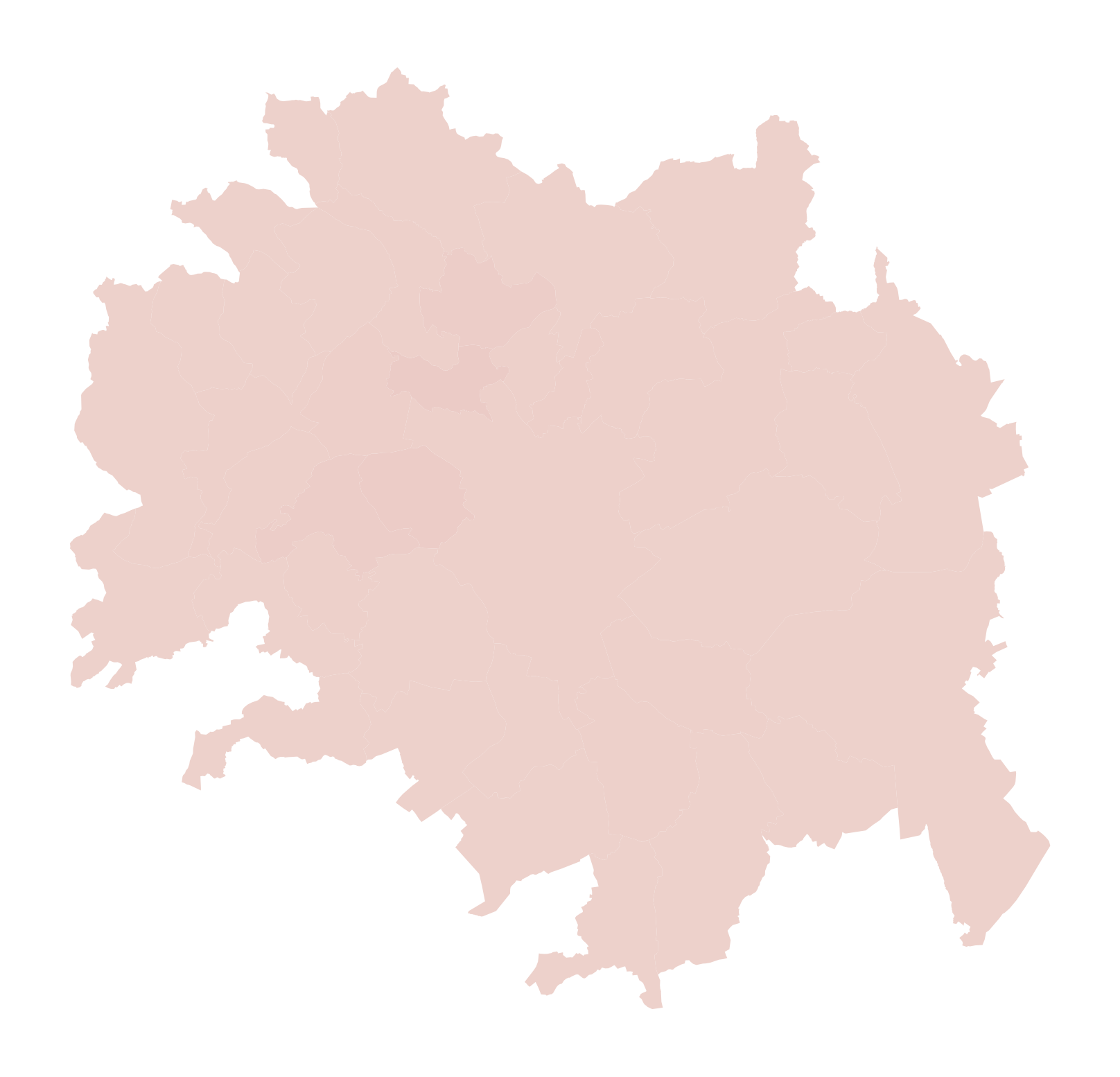}
         \caption{Le Mans}
         \label{fig:cpc_Mans}
     \end{subfigure}
     \begin{subfigure}[c]{0.18\linewidth}
         \centering
         \includegraphics[width=1\textwidth]{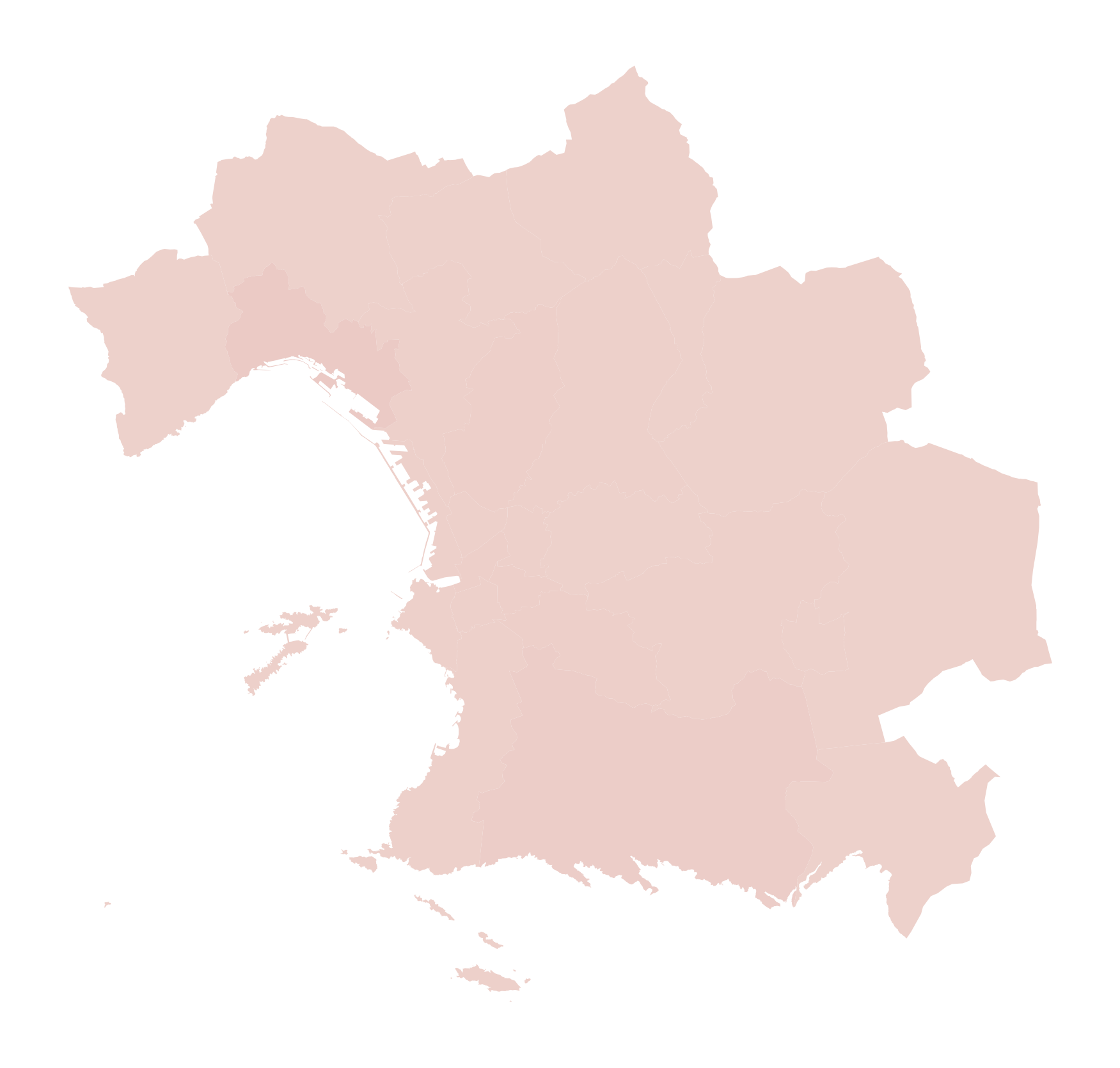}
         \caption{Marseille}
         \label{fig:cpc_Marseille}
     \end{subfigure}
     \hfill
     \begin{subfigure}[c]{0.18\linewidth}
         \centering
         \includegraphics[width=1\textwidth]{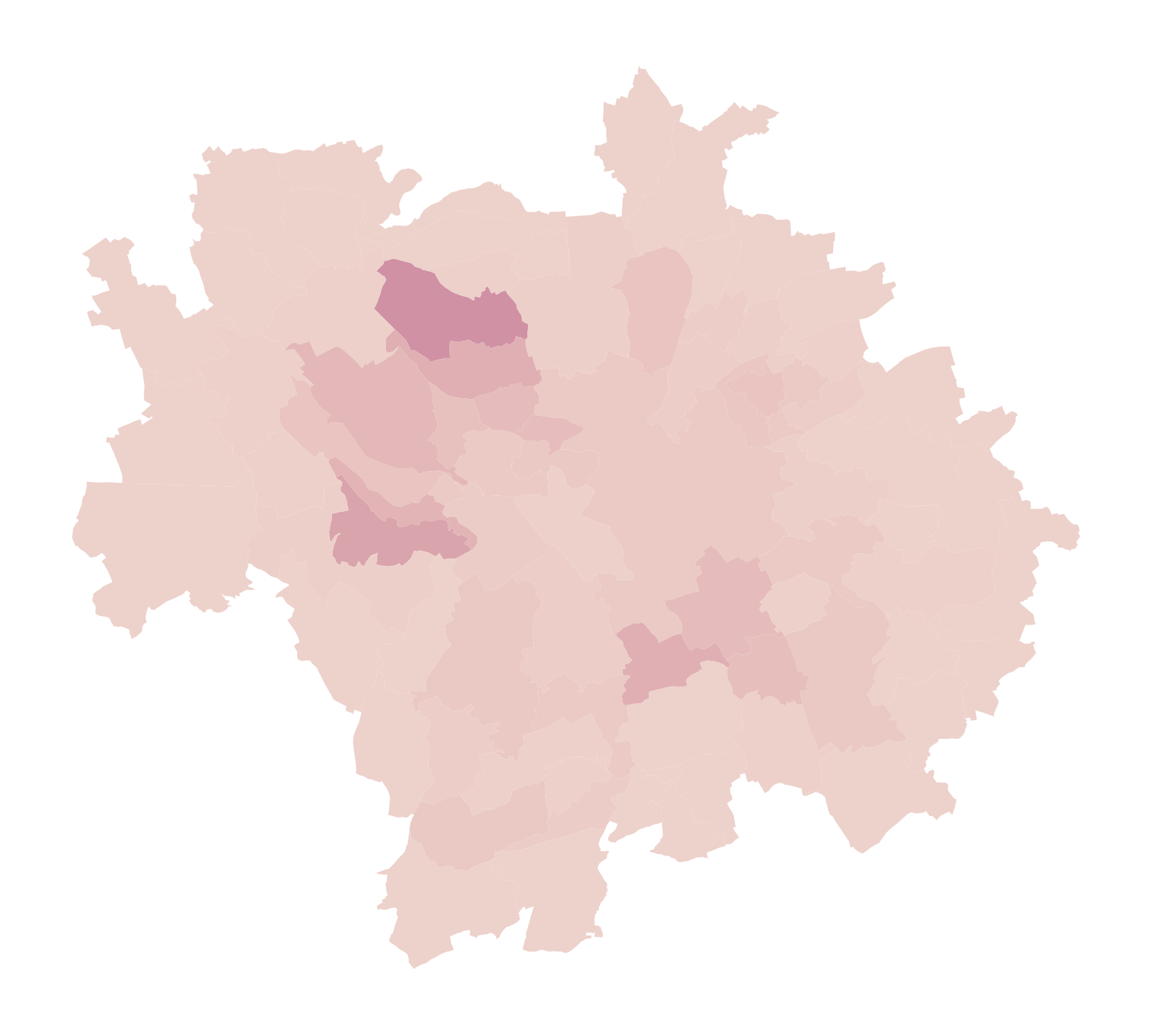}
         \caption{Metz}
         \label{fig:cpc_Metz}
     \end{subfigure}
     \begin{subfigure}[c]{0.18\linewidth}
         \centering
         \includegraphics[width=1\textwidth]{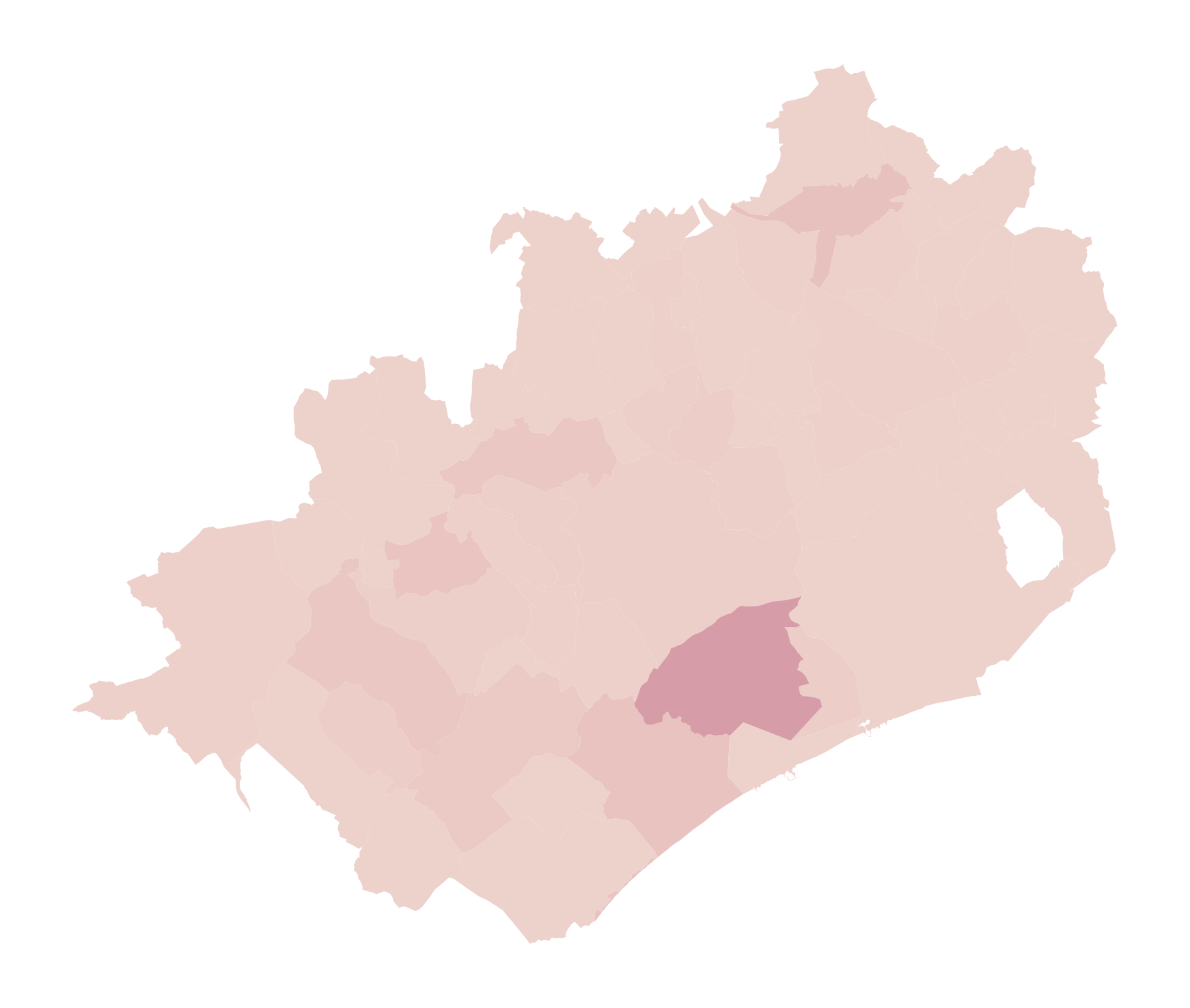}
         \caption{Montpellier}
         \label{fig:cpc_Montpellier}
     \end{subfigure}
     \begin{subfigure}[c]{0.18\linewidth}
         \centering
         \includegraphics[width=1\textwidth]{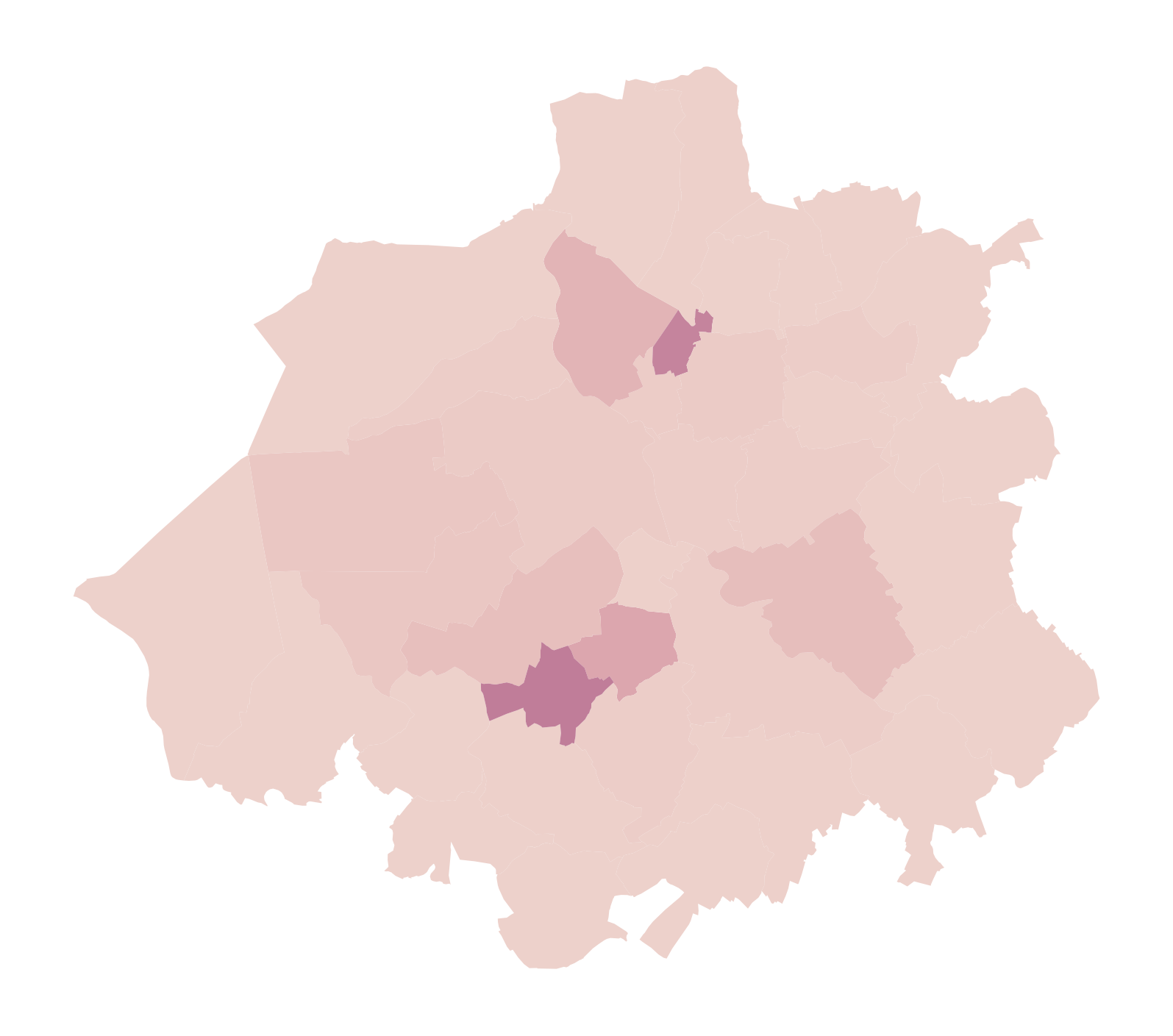}
         \caption{Nancy}
         \label{fig:cpc_Nancy}
     \end{subfigure}
     \begin{subfigure}[c]{0.18\linewidth}
         \centering
         \includegraphics[width=1\textwidth]{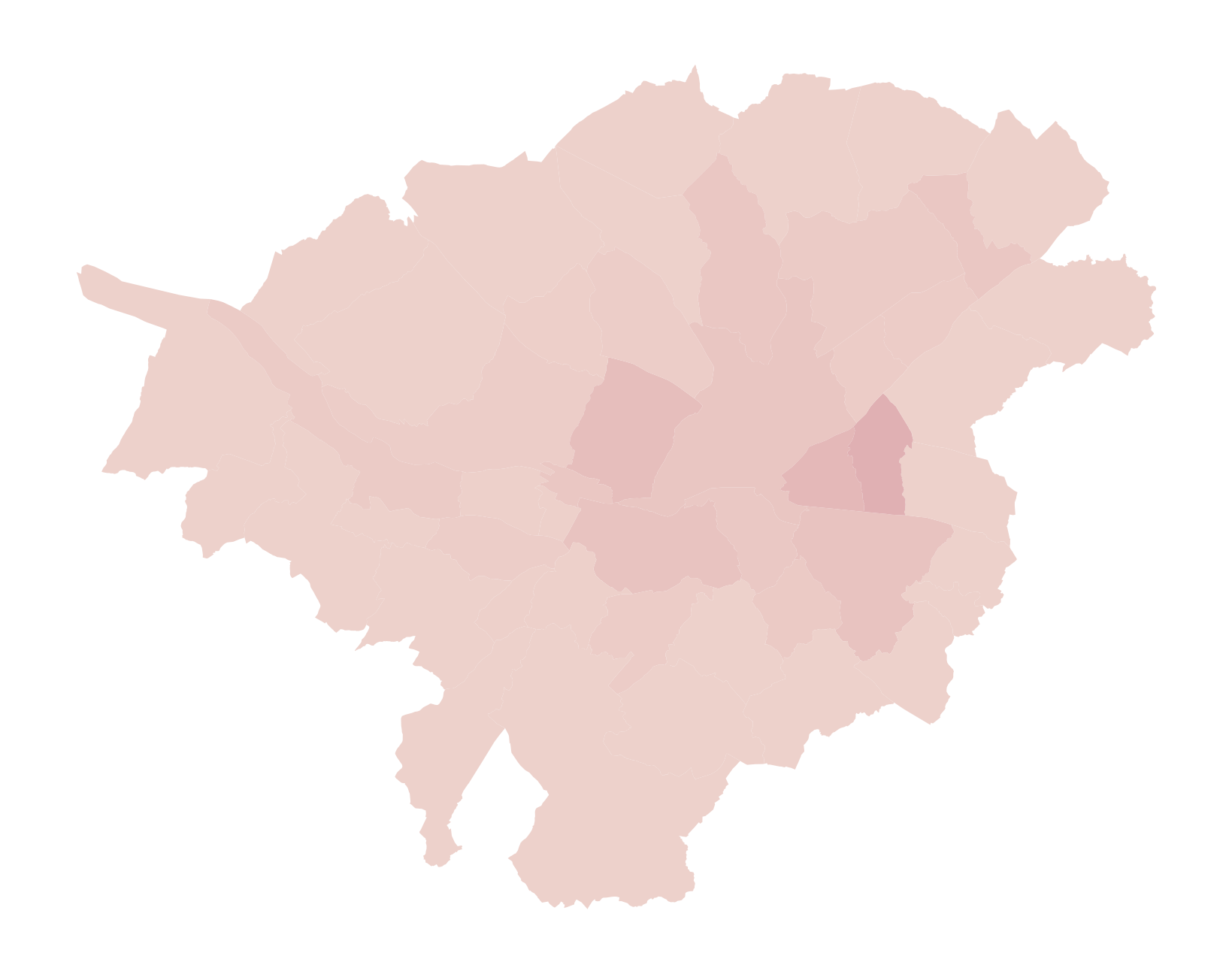}
         \caption{Nantes}
         \label{fig:cpc_Nantes}
     \end{subfigure}
     \hfill
     \begin{subfigure}[c]{0.18\linewidth}
         \centering
         \includegraphics[width=1\textwidth]{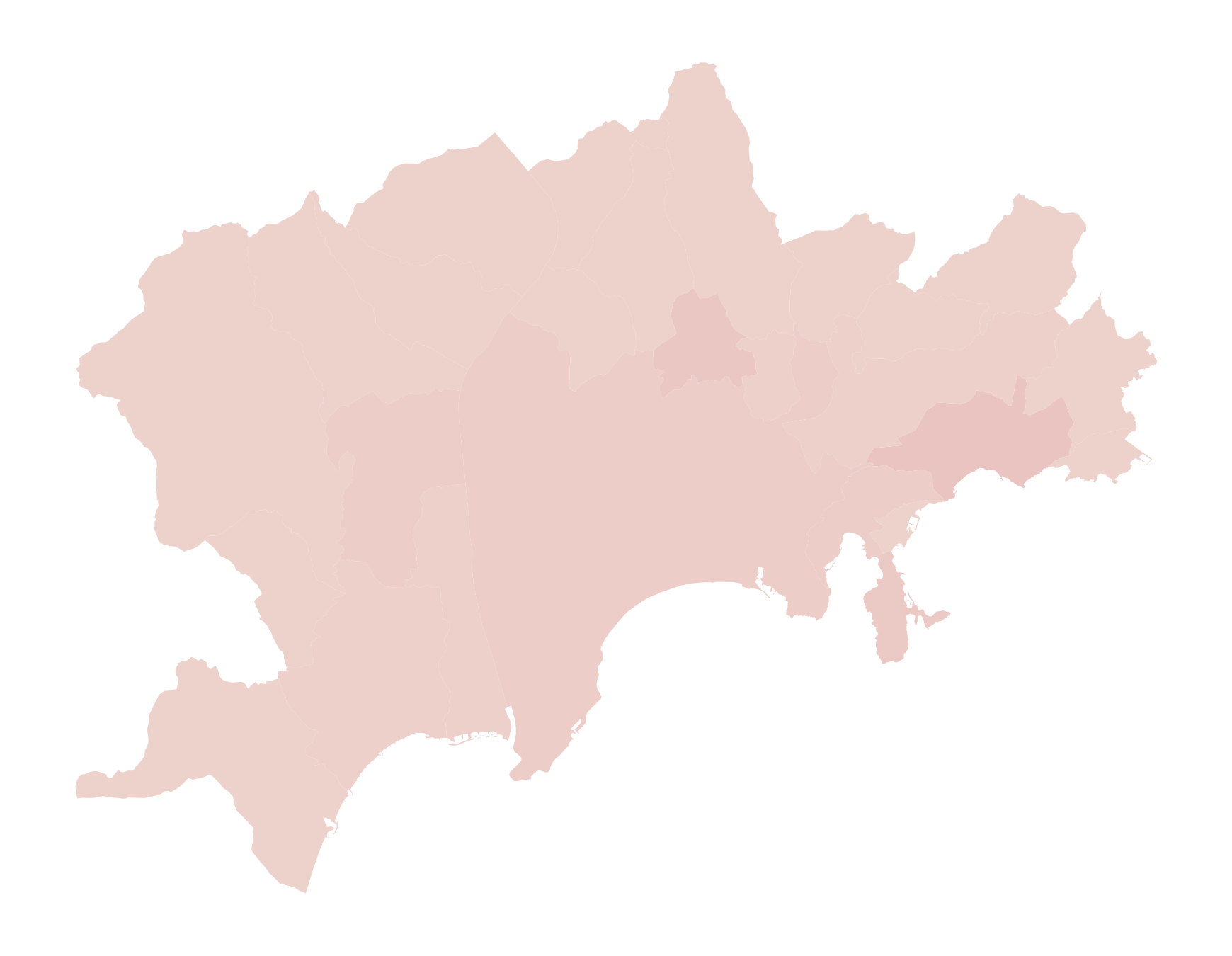}
         \caption{Nice}
         \label{fig:cpc_Nice}
     \end{subfigure}
     \begin{subfigure}[c]{0.18\linewidth}
         \centering
         \includegraphics[width=1\textwidth]{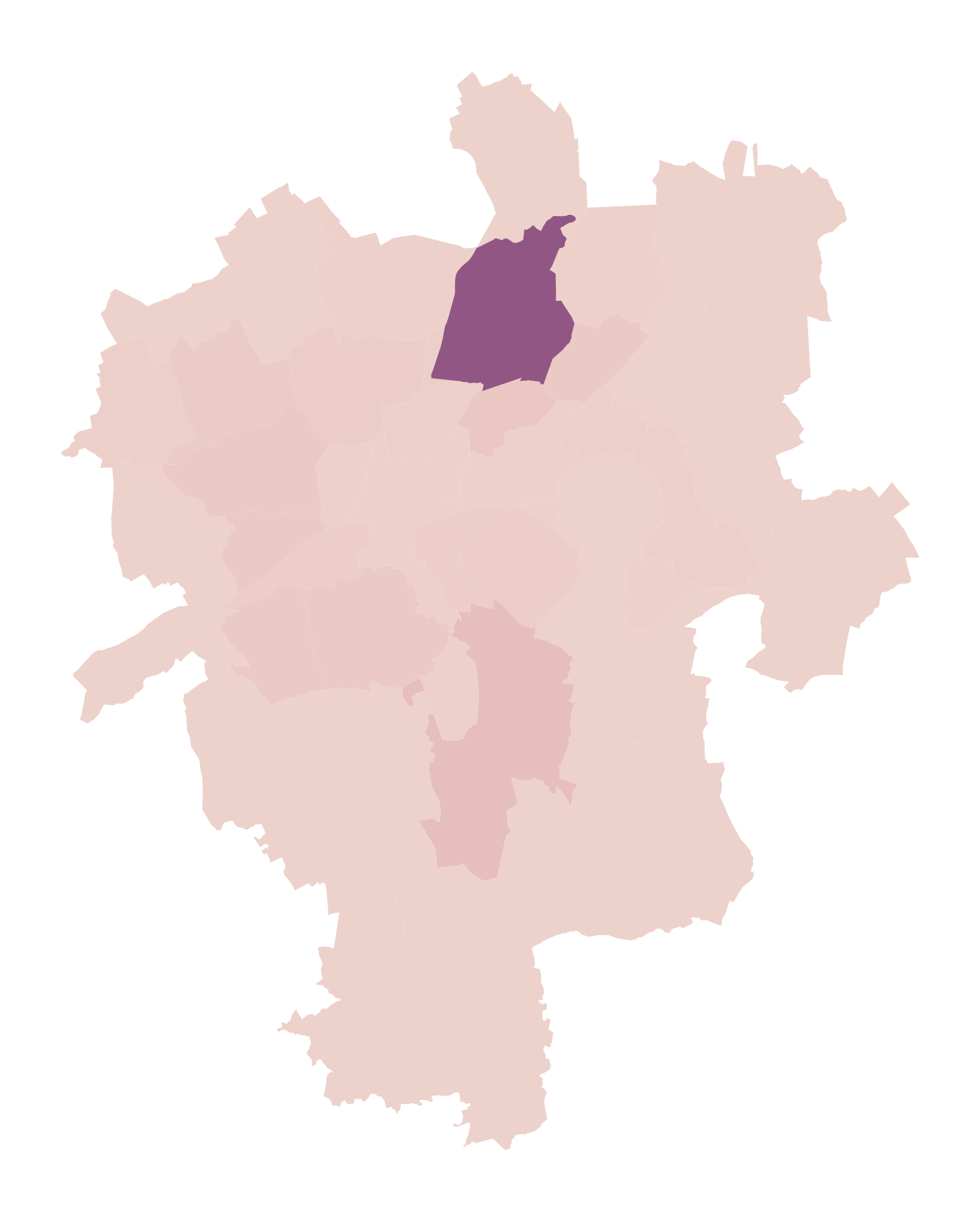}
         \caption{Orléans}
         \label{fig:cpc_Orleans}
     \end{subfigure}
     \begin{subfigure}[c]{0.18\linewidth}
         \centering
         \includegraphics[width=1\textwidth]{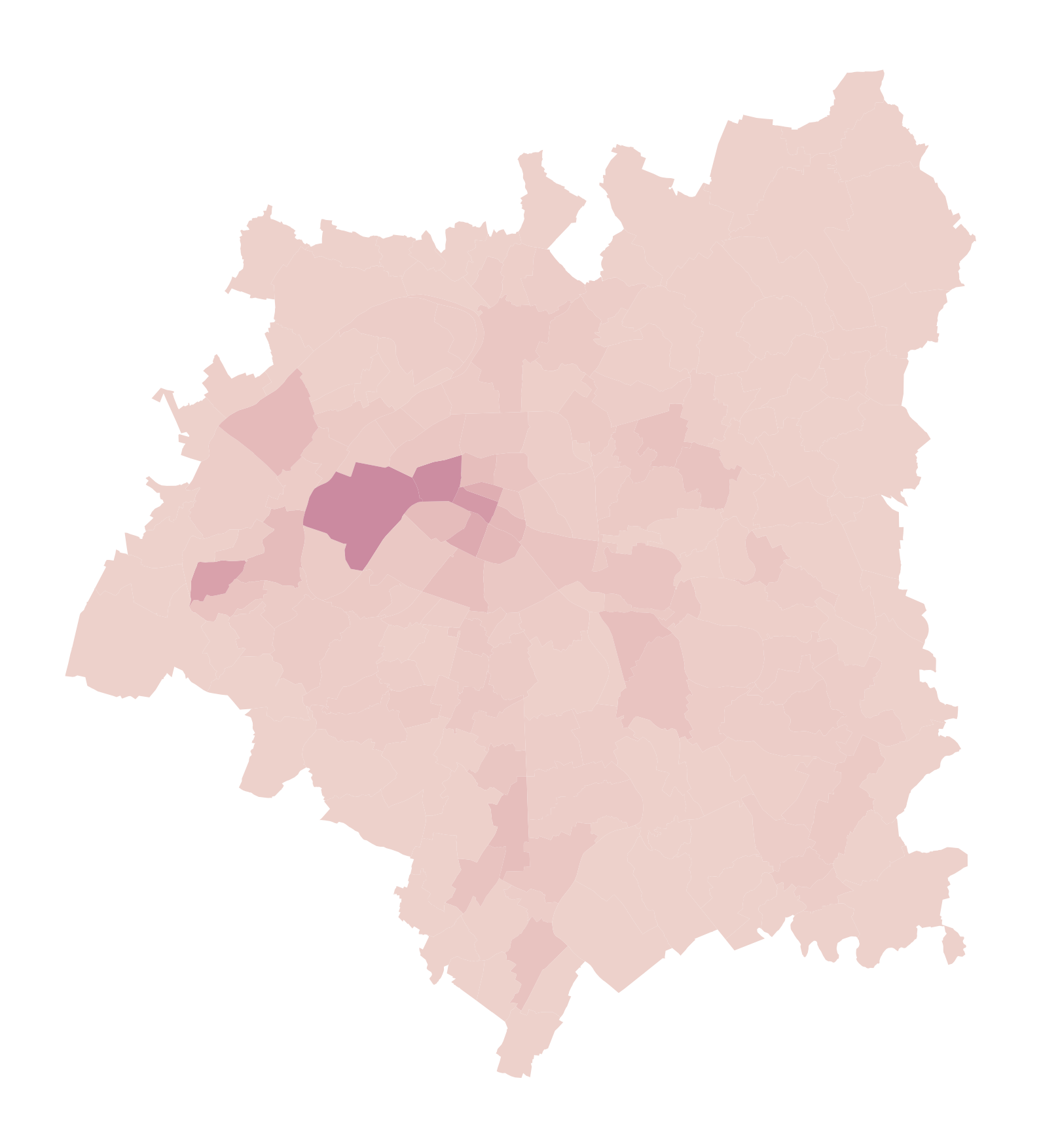}
         \caption{Paris}
         \label{fig:cpc_Paris}
     \end{subfigure}
     \begin{subfigure}[c]{0.18\linewidth}
         \centering
         \includegraphics[width=1\textwidth]{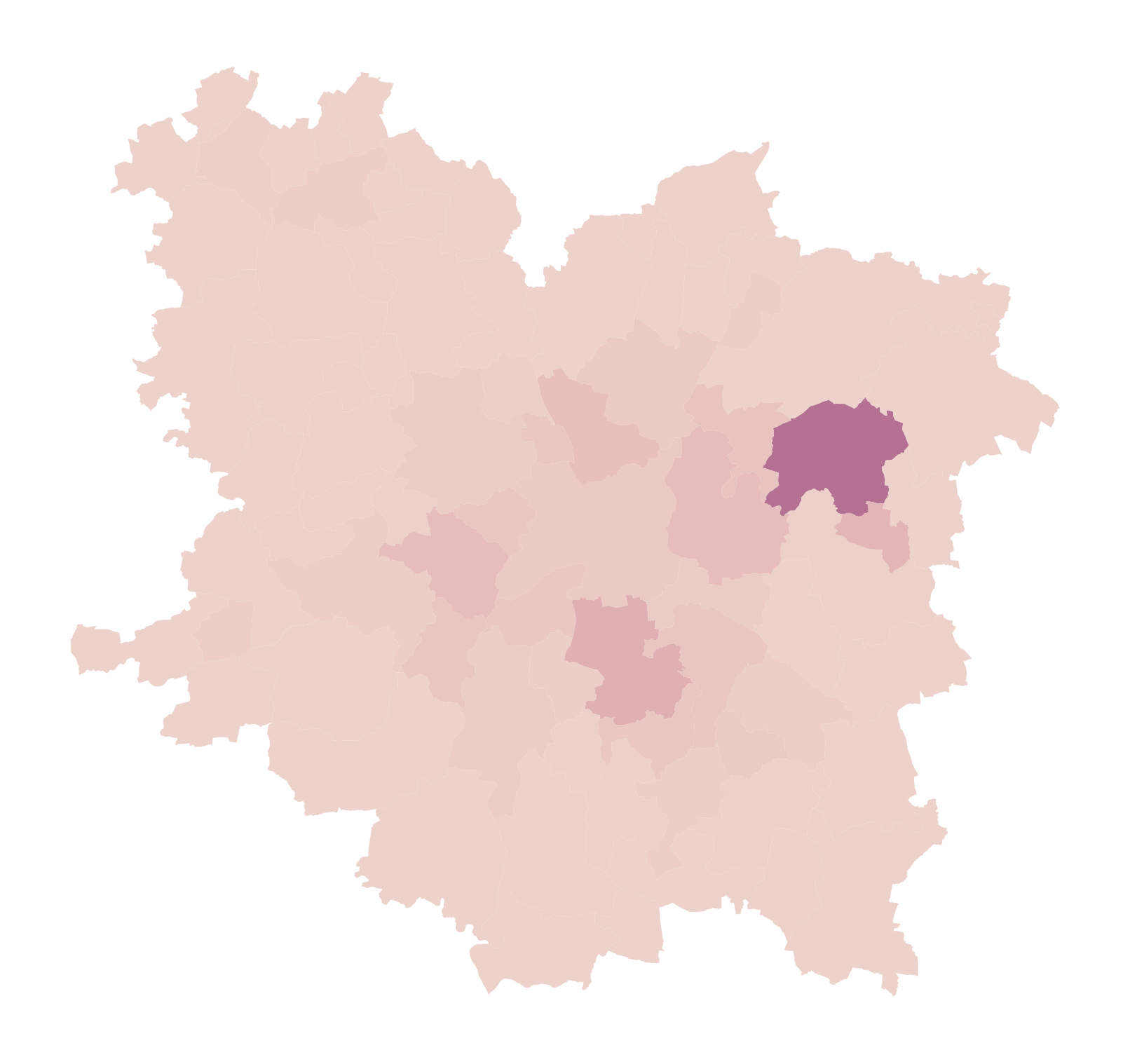}
         \caption{Rennes}
         \label{fig:cpc_Rennes}
     \end{subfigure}
     \hfill
     \begin{subfigure}[c]{0.18\linewidth}
         \centering
         \includegraphics[width=1\textwidth]{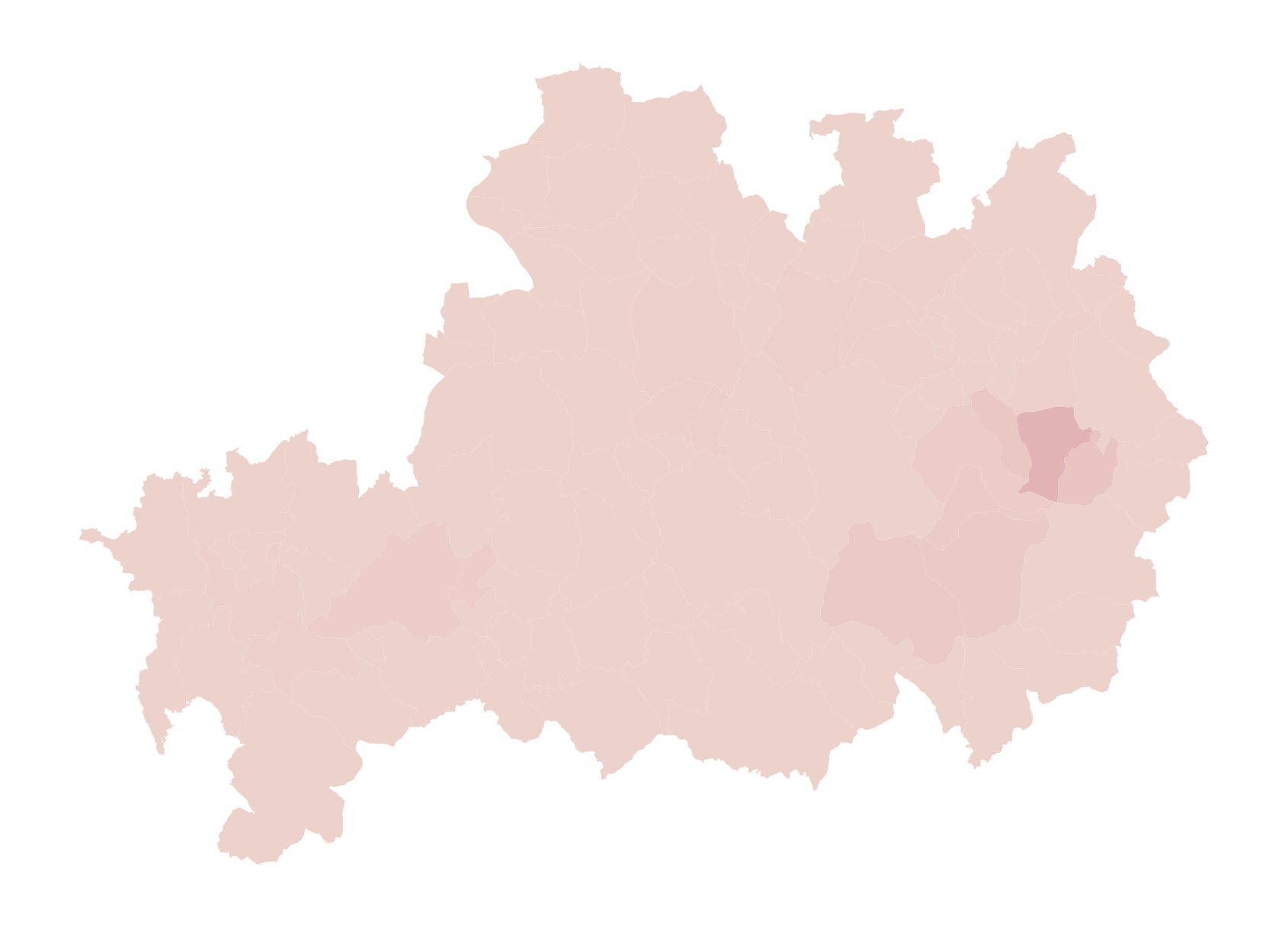}
         \caption{Saint-Étienne}
         \label{fig:cpc_Saint-Etienne}
     \end{subfigure}
     \begin{subfigure}[c]{0.18\linewidth}
         \centering
         \includegraphics[width=1\textwidth]{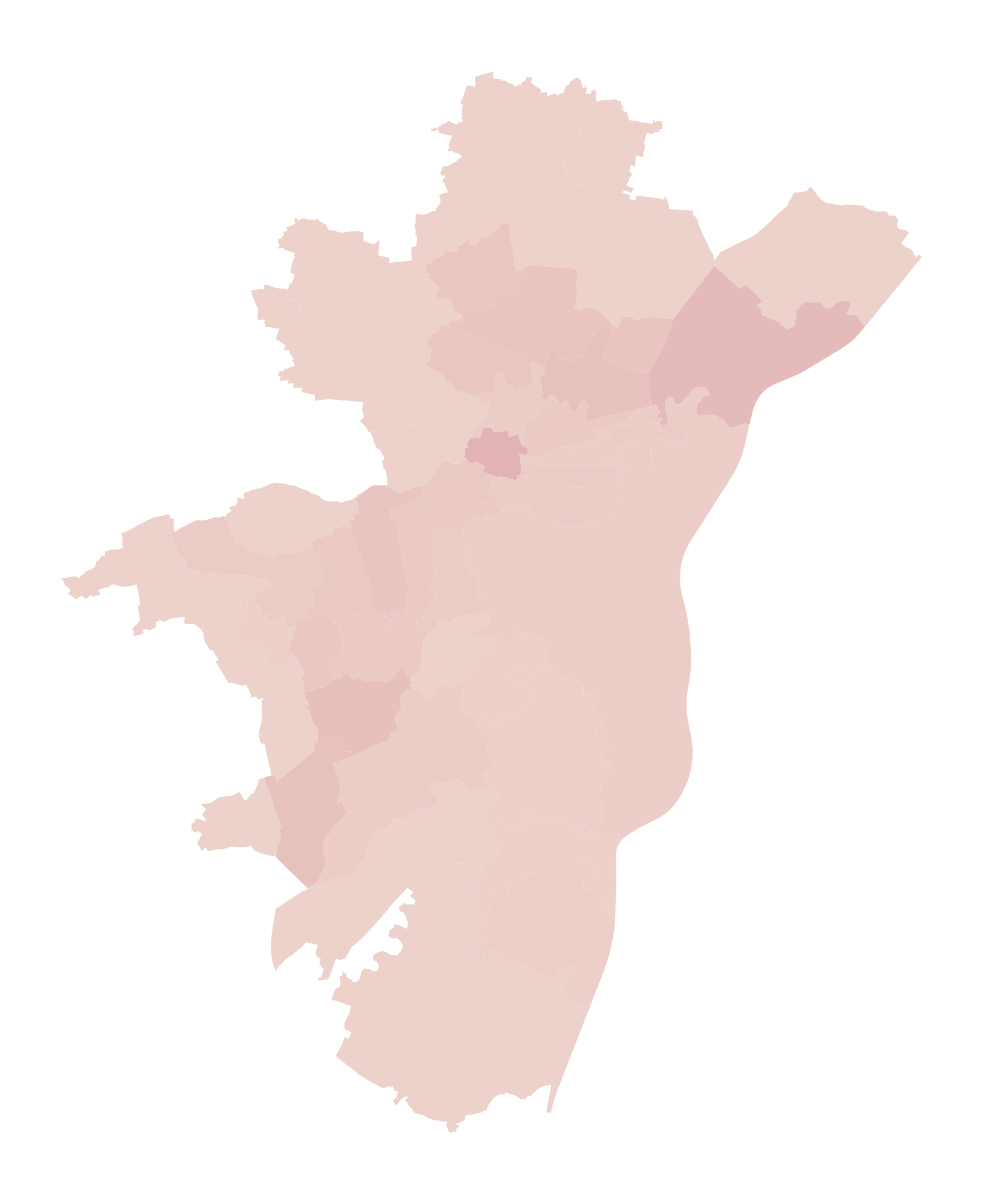}
         \caption{Strasbourg}
         \label{fig:cpc_Strasbourg}
     \end{subfigure}
     \begin{subfigure}[c]{0.18\linewidth}
         \centering
         \includegraphics[width=1\textwidth]{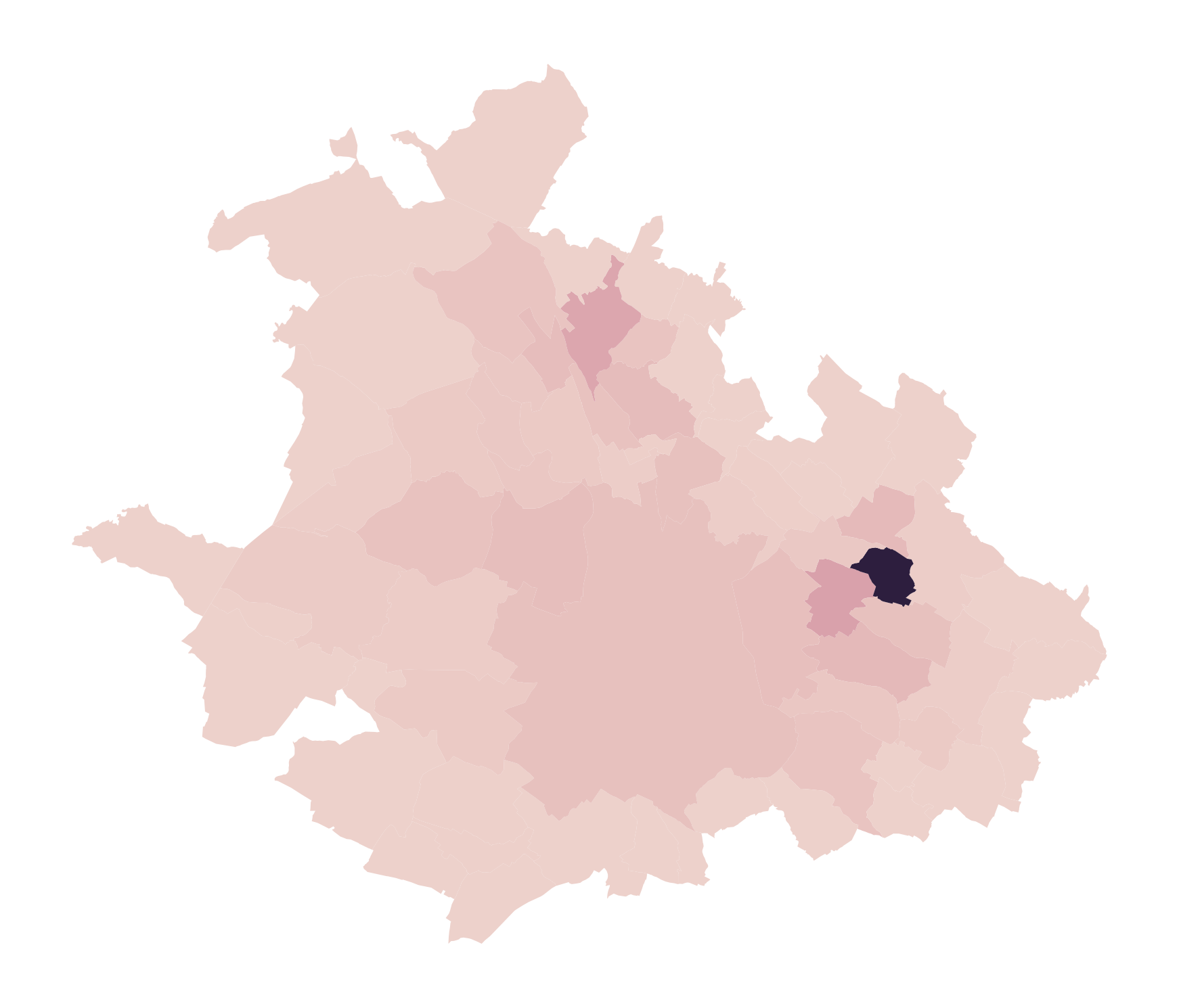}
         \caption{Toulouse}
         \label{fig:cpc_Toulouse}
     \end{subfigure}
     \begin{subfigure}[c]{0.18\linewidth}
         \centering
         \includegraphics[width=1\textwidth]{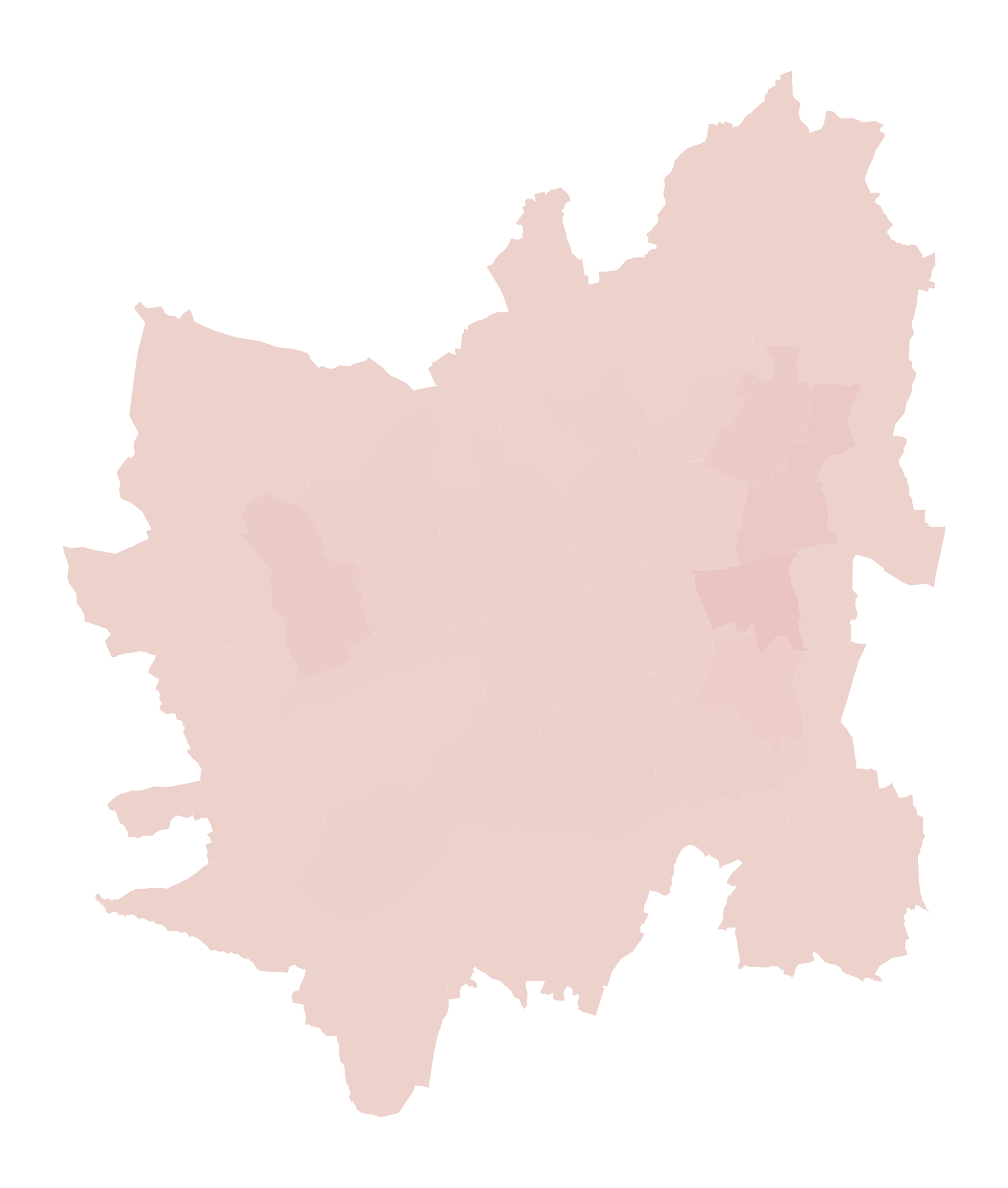}
         \caption{Tours}
         \label{fig:cpc_Tours}
     \end{subfigure}
     \hfill
     \begin{subfigure}[c]{1\linewidth}
         \centering
         \includegraphics[width=\textwidth]{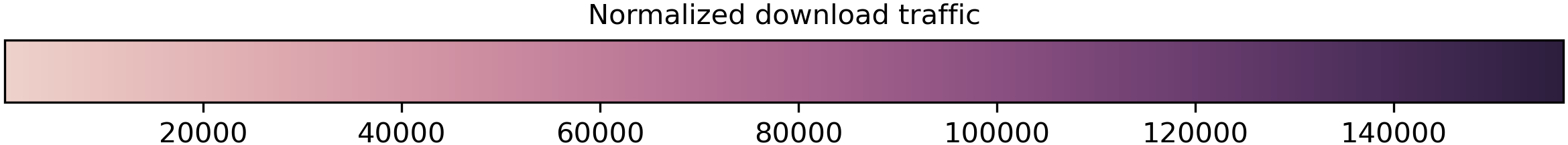}
     \end{subfigure}
     \label{fig:cpc_main}
\end{figure}

\newpage

\begin{table}[htbp]
\centering
\small
\caption{\textbf{Regression results for our cpc estimates for all communes controlling for the drug abuse rate, by commune.}} 
  \label{tab:robust-drug} 
\begin{tabular}{@{\extracolsep{5pt}}lcc} 
\\[-1.8ex]\hline 
\hline \\[-1.8ex] 
 & \multicolumn{2}{c}{\textit{Dependent variable (per 1000):}} \\ 
\cline{2-3} 
\\[-1.8ex] & \multicolumn{2}{c}{log\_cpc} \\ 
\hline \\[-1.8ex]  
 log\_YouTube\_per\_1000 & 0.32$^{**}$ & 0.32$^{**}$ \\ 
  & (0.14) & (0.14) \\ 
 log\_Web\_Adult\_per\_1000 & $-$0.24$^{*}$ & $-$0.24$^{*}$ \\ 
  & (0.13) & (0.13) \\ 
 log\_Tor\_per\_1000 & 1.14$^{***}$ & 1.14$^{***}$ \\ 
  & (0.05) & (0.05) \\ 
 log\_pop\_density & 0.15$^{***}$ & 0.15$^{***}$ \\ 
  & (0.05) & (0.05) \\ 
 Share\_of\_singles & 0.82 & 0.82 \\ 
  & (0.90) & (0.90) \\ 
 Drug\_abuse\_rate &  & $-$0.03 \\ 
  &  & (0.52) \\ 
 Poverty\_rate & $-$4.25$^{***}$ & $-$4.25$^{***}$ \\ 
  & (1.53) & (1.54) \\ 
 Employment\_rate & $-$8.50$^{***}$ & $-$8.48$^{***}$ \\ 
  & (2.54) & (2.60) \\ 
 Electoral\_turnout\_2017 & $-$2.35$^{*}$ & $-$2.36$^{*}$ \\ 
  & (1.26) & (1.25) \\ 
 Share\_Le\_Pen & $-$0.30 & $-$0.30 \\ 
  & (1.52) & (1.51) \\ 
 Share\_Macron & 2.10 & 2.10 \\ 
  & (2.18) & (2.19) \\ 
 POI\_adult\_entertainment & 0.003 & 0.003 \\ 
  & (0.01) & (0.01) \\ 
 POI\_sports\_teams & $-$0.004 & $-$0.005 \\ 
  & (0.02) & (0.02) \\ 
 POI\_church & $-$0.02 & $-$0.02 \\ 
  & (0.03) & (0.03) \\ 
 POI\_mosque & $-$0.04$^{**}$ & $-$0.04$^{**}$ \\ 
  & (0.02) & (0.02) \\ 
 POI\_religious\_org & $-$0.10$^{***}$ & $-$0.10$^{***}$ \\ 
  & (0.03) & (0.03) \\ 
 POI\_school & 0.01 & 0.01 \\ 
  & (0.02) & (0.02) \\ 
 PC1 & $-$0.05$^{**}$ & $-$0.05$^{**}$ \\ 
  & (0.02) & (0.02) \\ 
 PC3 & 0.03 & 0.03 \\ 
  & (0.05) & (0.05) \\ 
 Sexual\_violence\_per\_1000 & 0.14 & 0.14 \\ 
  & (0.13) & (0.16) \\ 
 Constant & $-$10.43$^{***}$ & $-$10.43$^{***}$ \\ 
  & (1.24) & (1.24) \\ 
\hline \\[-1.8ex]  
Observations & 430 & 430 \\ 
R$^{2}$ & 0.96 & 0.96 \\ 
Adjusted R$^{2}$ & 0.96 & 0.96 \\ 
\hline 
\hline \\[-1.8ex] 
\textit{Note: Heteroscedasticity-robust SE}  & \multicolumn{2}{r}{$^{*}$p$<$0.1; $^{**}$p$<$0.05; $^{***}$p$<$0.01} \\ 
\end{tabular} 
\end{table}

\newpage

\begin{table}[htbp] \centering
\small
\caption{\textbf{Regression results without zero reported cases, by commune.}} 
  \label{tab:robust-nozero} 
\begin{tabular}{@{\extracolsep{5pt}}lcc} 
\\[-1.8ex]\hline 
\hline \\[-1.8ex] 
 & \multicolumn{2}{c}{\textit{Dependent variable (per 1000):}} \\ 
\cline{2-3} 
\\[-1.8ex] & log\_cpc & Sexual\_violence\\ 
\hline \\[-1.8ex]  
 log\_YouTube\_per\_1000 & 0.07 & 0.28 \\ 
  & (0.08) & (0.26) \\ 
 log\_Web\_Adult\_per\_1000 & 0.08 & $-$0.34 \\ 
  & (0.08) & (0.30) \\ 
 log\_Tor\_per\_1000 & 1.00$^{***}$ & $-$0.20 \\ 
  & (0.03) & (0.24) \\ 
 log\_pop\_density & 0.09$^{***}$ & $-$0.30 \\ 
  & (0.03) & (0.28) \\ 
 Share\_of\_singles & $-$0.08 & 6.23 \\ 
  & (0.50) & (4.15) \\ 
 Poverty\_rate & $-$2.92$^{**}$ & 2.52 \\ 
  & (1.44) & (4.09) \\ 
 Employment\_rate & $-$5.57$^{***}$ & 2.75 \\ 
  & (2.09) & (2.32) \\ 
 Electoral\_turnout\_2017 & 0.34 & 0.81 \\ 
  & (0.65) & (0.74) \\ 
 Share\_Le\_Pen & $-$1.06 & 3.15 \\ 
  & (0.89) & (2.59) \\ 
 Share\_Macron & $-$0.68 & 0.32 \\ 
  & (1.01) & (2.21) \\ 
 POI\_adult\_entertainment & 0.01 & $-$0.01 \\ 
  & (0.01) & (0.03) \\ 
 POI\_sports\_teams & 0.01 & 0.01 \\ 
  & (0.02) & (0.04) \\ 
 POI\_church & $-$0.10$^{*}$ & $-$0.04 \\ 
  & (0.06) & (0.13) \\ 
 POI\_mosque & $-$0.0001 & $-$0.005 \\ 
  & (0.01) & (0.02) \\ 
 POI\_religious\_org & $-$0.03$^{*}$ & 0.01 \\ 
  & (0.02) & (0.03) \\ 
 POI\_school & $-$0.03 & 0.15 \\ 
  & (0.03) & (0.10) \\ 
 PC1 & 0.01 & 0.04 \\ 
  & (0.02) & (0.05) \\ 
 PC3 & $-$0.10$^{**}$ & 0.06 \\ 
  & (0.05) & (0.08) \\ 
 Sexual\_violence\_per\_1000 & 0.07 &  \\ 
  & (0.12) &  \\ 
 log\_cpc\_per\_1000 &  & 0.21 \\ 
  &  & (0.25) \\ 
 Constant & $-$9.82$^{***}$ & 0.54 \\ 
  & (0.69) & (1.33) \\ 
\hline \\[-1.8ex]  
Observations & 326 & 326 \\ 
R$^{2}$ & 0.98 & 0.27 \\ 
Adjusted R$^{2}$ & 0.98 & 0.22 \\ 
\hline 
\hline \\[-1.8ex] 
\textit{Note: Heteroscedasticity-robust SE}  & \multicolumn{2}{r}{$^{*}$p$<$0.1; $^{**}$p$<$0.05; $^{***}$p$<$0.01} \\ 
\end{tabular} 
\end{table}

\newpage

\begin{table}[htbp] \centering
\small
\caption{\textbf{Regression results for our cpc estimates with (1) and without (2) a two-hour time lag of the Tor traffic, by commune.}} 
  \label{tab:robust-lag} 
\begin{tabular}{@{\extracolsep{5pt}}lcc} 
\\[-1.8ex]\hline 
\hline \\[-1.8ex] 
 & \multicolumn{2}{c}{\textit{Dependent variable (per 1000):}} \\ 
\cline{2-3} 
\\[-1.8ex] & \multicolumn{2}{c}{log\_cpc} \\ 
\\[-1.8ex] & no lag & lagged\\ 
\hline \\[-1.8ex]  
 log\_YouTube\_per\_1000 & 0.35$^{***}$ & $-$0.06 \\ 
  & (0.11) & (0.17) \\ 
 log\_Web\_Adult\_per\_1000 & $-$0.25$^{**}$ & 0.13 \\ 
  & (0.10) & (0.13) \\ 
 log\_Tor\_per\_1000 & 1.13$^{***}$ & 1.14$^{***}$ \\ 
  & (0.04) & (0.06) \\ 
 log\_pop\_density & 0.14$^{***}$ & 0.23$^{***}$ \\ 
  & (0.03) & (0.04) \\ 
 Share\_of\_singles & 0.80 & 1.06 \\ 
  & (0.66) & (0.96) \\ 
 Poverty\_rate & $-$5.21$^{***}$ & $-$7.01$^{***}$ \\ 
  & (1.17) & (1.77) \\ 
 Employment\_rate & $-$8.40$^{***}$ & $-$6.25$^{*}$ \\ 
  & (2.24) & (3.54) \\ 
 Electoral\_turnout\_2017 & $-$2.27$^{**}$ & 0.76 \\ 
  & (0.96) & (1.26) \\ 
 Share\_Le\_Pen & 0.59 & $-$3.40$^{***}$ \\ 
  & (1.02) & (1.31) \\ 
 Share\_Macron & 2.72$^{*}$ & $-$2.14 \\ 
  & (1.54) & (1.74) \\ 
 POI\_adult\_entertainment & $-$0.003 & 0.01 \\ 
  & (0.01) & (0.03) \\ 
 POI\_sports\_teams & $-$0.001 & 0.03$^{*}$ \\ 
  & (0.01) & (0.02) \\ 
 POI\_church & $-$0.01 & 0.03 \\ 
  & (0.01) & (0.03) \\ 
 POI\_mosque & $-$0.03$^{**}$ & $-$0.01 \\ 
  & (0.01) & (0.03) \\ 
 POI\_religious\_org & $-$0.06$^{*}$ & $-$0.05 \\ 
  & (0.04) & (0.05) \\ 
 POI\_school & $-$0.02 & $-$0.02 \\ 
  & (0.02) & (0.03) \\ 
 PC1 & $-$0.03$^{*}$ & $-$0.02 \\ 
  & (0.02) & (0.03) \\ 
  & & \\ 
 PC3 & $-$0.01 & $-$0.17$^{***}$ \\ 
  & (0.03) & (0.05) \\ 
 Sexual\_violence\_per\_1000 & 0.14$^{**}$ & 0.23$^{***}$ \\ 
  & (0.07) & (0.06) \\ 
 Constant & $-$11.04$^{***}$ & $-$10.73$^{***}$ \\ 
  & (1.00) & (1.47) \\ 
\hline \\[-1.8ex]  
Observations & 630 & 630 \\ 
R$^{2}$ & 0.96 & 0.92 \\ 
Adjusted R$^{2}$ & 0.96 & 0.92 \\ 
\hline 
\hline \\[-1.8ex] 
\textit{Note: Heteroscedasticity-robust SE}  & \multicolumn{2}{r}{$^{*}$p$<$0.1; $^{**}$p$<$0.05; $^{***}$p$<$0.01} \\ 
\end{tabular} 
\end{table}



\end{document}